\documentclass[11pt,letterpaper]{article}
\usepackage[latin1]{inputenc}
\usepackage{latexsym}
\usepackage{amssymb}
\usepackage{amsmath}
\usepackage{graphicx}
\usepackage{amsfonts}
\usepackage[nottoc]{tocbibind}

\newcommand{\onetothree}{(x^{1},\,x^{2},\,x^{3})}
\newcommand{\onetofour}{(x^{1},\,x^{2},\,x^{3},\,x^{4})}
\newcommand{\Chi}{\mathcal{X}}
\newcommand{\relgamma}{\sqrt{1-\left(\frac{v^{1}}{c}\right)^{2}}}
\newcommand{\genrelgammax}{\sqrt{1-\frac{|\mathbf{V}(x)|^{2}}{c^{2}}}}
\newcommand{\genrelgamma}{\sqrt{1-\frac{|\mathbf{V}(t)|^{2}}{c^{2}}}}
\newcommand{\flgenrelgamma}{\sqrt{1-\frac{|\mathbf{v}|^{2}}{c^{2}}}}
\newcommand{\genrelgammaxt}{\sqrt{1-\frac{|\mathbf{V}(\mathbf{x},t)|^{2}}{c^{2}}}}
\newcommand{\bfx}{\mathbf{x}}
\newcommand{\plusfact}{1+\frac{2|W(\mathbf{x})|}{c^{2}}}
\newcommand{\minusfact}{1-\frac{2|W(\mathbf{x})|}{c^{2}}}
\newcommand{\plusfactnabs}{1+\frac{2W(\mathbf{x})}{c^{2}}}
\newcommand{\minusfactnabs}{1-\frac{2W(\mathbf{x})}{c^{2}}}

\begin{document}
\title{Relativistic particle, fluid and plasma mechanics coupled to gravity}
\author {{\small A. Das \footnote{e-mail: das@sfu.ca}} \\
\it{\small Department of Mathematics} \\ \it{\small Simon Fraser
University, Burnaby, British Columbia, Canada V5A 1S6}
 \and
{\small A. DeBenedictis \footnote{e-mail: adebened@sfu.ca}} \\
\it{\small Department of Physics} \\ \it{\small Simon Fraser
University, Burnaby, British Columbia, Canada V5A 1S6}
\and
{\small S. Kloster \footnote{e-mail: stevek@sfu.ca}} \\
\it{\small Centre for Experimental and Constructive Mathematics} \\ \it{\small Simon Fraser
University, Burnaby, British Columbia, Canada V5A 1S6}
\and
{\small N. Tariq \footnote{e-mail: ntariq@sfu.ca}} \\
\it{\small Department of Mathematics} \\ \it{\small Simon Fraser
University, Burnaby, British Columbia, Canada V5A 1S6}
}
\date{{\small March 10, 2005}}
\maketitle

\begin{abstract}
\noindent In this introductory review article, we explore the
special relativistic equations of particle  motions and the
consequent derivation of Einstein's famous formula $E=mc^2$.
Next, we study the special relativistic electromagnetic field
equations and generalizations of Lorentz equations of motion for
charged particles. We then introduce the special relativistic
gravitational field as a symmetric second order tensor field.
Particle motions in the presence of static gravity are explored
which could be used to study planetary dynamics, revealing
perihelion shifts. Next, we investigate the system of consisting
of pressureless plasmas and neutral perfect fluids coupled to the
gravitational field. In that arena, we derive the relativistic
Euler equation. Finally, we investigate the relativistic dynamics
of a perfect fluid plasma and extensions to viscous flow and
derive the relativistic Navier-Stokes equation.
\end{abstract}

\vspace{3mm}
\newpage
\tableofcontents
\newpage

\section{Introduction}
A century has elapsed since the momentous discovery of the special
theory of relativity by Einstein \cite{ref:1}. This theory has
inspired subsequent pursuit of the general theory of relativity
as a novel theory of gravitation \cite{ref:2}. It has also led to
the understanding of the relativistic wave equation of an
electron \cite{ref:3}. In the last five decades, relativity has
also inspired the gauge field theories of subatomic particle
interactions which have spectacular experimental confirmations
\cite{ref:4}. However, applications of relativity theory into
applied mathematical problems are almost non-existent. We venture
to write this review article mainly to attract the academic
attentions of applied mathematicians to this fascinating branch
of modern theoretical science.

\qquad There exist well established special relativistic particle
mechanics, relativistic fluid mechanics and electrodynamics. In
the special theory of relativity, spacetime is assumed to be a
\emph{flat} differentiable manifold. Since Einstein's
gravitational theory involves a \emph{curved} pseudo-Riemannian
manifold, special relativistic dynamics of various macroscopic
systems usually must omit gravitation. This is generally valid as
the gravitational field is extremely weak in comparison to the
other force fields involved. However, a linearized version of
Einstein's theory of gravitation can be incorporated within the
framework of special relativity. Our present review article aims
at such a treatment of various dynamical systems. It is hoped
that such a review will serve as a useful introduction to the
field for practitioners of non relativistic fluid mechanics as
well as those wishing to study weak-field gravitating particles,
fluids and plasmas such as are found in various astrophysical
systems. As mentioned above, it is written with applied
mathematicians as the primary intended audience. However, we hope
this review will be useful to the wider audience as well. No
previous knowledge of special relativity or Minkowski tensors is
assumed.

\qquad The electrodynamics and mechanics of relativistic
continuous media, coupled to gravitation, has many interesting
applications in astrophysics (for examples, see \cite{ref:4b})
and other areas of general relativity and high-energy physics
\cite{ref:4c}. As well, the intrested reader is refered to the
books \cite{ref:4d}  and references therein.

\qquad In section-II, notations for vectors and tensors in three
and four dimensions are laid out in a leisurely fashion. In the
third section, particle mechanics (mainly special relativistic)
is discussed in a nut-shell. We do derive Einstein's famous
equation $E=mc^2$ in this section.

\qquad In section-IV, we discuss Maxwell's equations of
electromagnetic fields. These equations are known to be already
relativistic! However, Lorentz's equation of motion requires some
minor modification for the relativistic conversion.

\qquad In section-V, a special relativistic version of the
gravitational field equations is investigated. It involves a
symmetric second order tensor field in spacetime to represent the
gravitational force. We then couple the gravitational field
with: (i) an incoherent dust, (ii) an electrically charged dust
(pressureless plasma), and (iii) a perfect fluid. In the
following section we specialize furthermore to static gravitational fields.
Planetary motions are investigated in the static field and an approximation to the
famous perihelion shift is derived. Moreover, we
also explore a perfect fluid in the presence of static external
gravitation, deriving the relativistic Euler equation in the process.

\qquad In the last section, we generalize preceding investigations
to more complicated materials (plasmas with pressure and
viscosity) and derive the relativistic Navier-Stokes equation.
Furthermore, generalizations to curvilinear coordinates and
orthonormal (or physical) components are also touched upon.

\qquad In deformable media, a stress tensor,
$\sigma_{ij}(x^{1},\,x^{2},\,x^{3})\equiv
\sigma_{ji}(x^{1},\,x^{2},\,x^{3})$ indicate equilibrium
conditions by the satisfaction of the equations
$\sum_{j=1}^{3}\frac{\partial\,\sigma_{ij}(x^{1},\,x^{2},\,x^{3})}{\partial
x^{j}}=0$. In the special relativistic generalization, an
energy-momentum-stress tensor
$T_{\mu\nu}(x^{1},\,x^{2},\,x^{3},\,x^{4})\equiv
T_{\nu\mu}(x^{1},\,x^{2},\,x^{3},\,x^{4})$ is introduced.
Moreover, \emph{dynamical equilibrium} of deformable bodies or fluids
are characterized by conditions
$\sum_{\nu=1}^{4}\frac{\partial\,T_{\mu\nu}(x^{1},\,x^{2},\,x^{3},\,x^{4})}{\partial
x^{\nu}}=0$. These equations are of utmost importnace throughout
the paper in deriving relativistic continuity equations and
relativistic equations for streamlines.

\section{Notations and conventions}
The three-dimensional physical space is assumed to be Euclidean. A
typical point in this space is denoted by
$\mathbf{x}:=\onetothree \in \mathbb{R}^{3}$. We use mostly a
Cartesian coordinate system.

\qquad The spacetime continuum, $M_{4}$, is assumed to be a flat
differentiable manifold admitting Minkowskian coordinate systems
\cite{ref:5}\cite{ref:6}. (These are generalizations of Cartesian
coordinates.) Relative to a Minkowski coordinate system, an
idealized point event in $M_{4}$ can be mapped uniquely into the
point $x:=\onetofour \in \mathbb{R}^{4}$. Here, $\onetothree$
indicates the spatial coordinates in the Euclidean
$\mathbb{R}^{3}$, whereas $x^{4}$ is the speed of light times the
time coordinate, i.e. $x^{4}=ct$.

\qquad Roman indices are used for three-dimensional spatial
components of vectors and tensors (the components in the
Euclidean $\mathbb{R}^{3}$). Greek indices are used for the four
dimensional spacetime components of vectors and tensors.
Einstein's summation convention is followed for \emph{both} Roman
and Greek indices.

\qquad We shall furnish some simple examples. three-dimensional
vectors are denoted by a \emph{bold} letter. For example, the
vector $\mathbf{v}$ has components $v^{i}\,;\, i\in
\left\{1,\,2,\,3\right\}$. The Kronecker delta:
$\delta^{i}_{\;j}$, $\delta_{ij}$, $\delta^{ij}$ are all entries
of the $3\times 3$ unit matrix $\left[I\right]_{3\times 3}$.
Therefore,
\begin{equation}
\delta^{i}_{\;j}\equiv \delta_{ij}\equiv \delta^{ij}:=\left\{
\begin{array}{lll}
1 & \mbox{ for }  & i=j, \\
0 & \mbox{ for }&
i \neq j.
\end{array}
\right. \label{eq:2.1}
\end{equation}
The inner product of two (three-dimensional) vectors
$\mathbf{v}$, $\mathbf{w}$ and the length $|\mathbf{v}|$ are given
by
\begin{subequations}
\begin{align}
\mathbf{v}\cdot\mathbf{w}=&\delta_{ij}v^{i}w^{j}=: \sum_{i=1}^{3}\sum_{j=1}^{3} \delta_{ij}v^{i}w^{j} \label{eq:2.2i} \\
|\mathbf{v}|:=&
+\sqrt{\mathbf{v}\cdot\mathbf{v}}=\sqrt{\delta_{ij}v^{i}v^{j}} =
\sqrt{(v^{1})^{2}+(v^{2})^{2} +(v^{3})^{2}}. \label{eq:2.2ii}
\end{align}
\end{subequations}
The totally anti-symmetric numerical (oriented) tensor \cite{ref:7} is defined by the components:
\begin{equation}
\epsilon_{ijk} :=\left\{
\begin{array}{lll}
+1 & \mbox{ for } (ijk) \;\;\mbox{an even permutation of }(123), \\
-1 & \mbox{ for } (ijk)  \;\;\mbox{an odd permutation of }(123), \\
\;\;\; 0 & \mbox{ otherwise}.
\end{array}
\right. \label{eq:2.3}
\end{equation}

We can express some familiar vector calculus notions with the help of $\epsilon_{ijk}$. For example,
\begin{subequations}
\begin{align}
\epsilon_{ijk}v^{j}w^{k}=&\left(\mathbf{v} \times \mathbf{w}\right)^{i} , \label{eq:2.4i}\\
\epsilon_{ijk} \frac{\partial B^{j}(x)}{\partial
x^{k}}=&-\left(\nabla \times \mathbf B\right)^{i}. \label{eq:2.4ii}
\end{align}
\end{subequations}
In the four-dimensional Minkowski spacetime, the metric tensor components are furnished by
\begin{equation}
d_{\mu\nu} \equiv d_{\nu\mu}:=\left\{
\begin{array}{lll}
\;\; \delta_{ij} & \mbox{ for }   \mu=i \mbox{ and } \nu=j, \\
-1 & \mbox{ for } \mu=\nu=4, \\
\;\;\; 0 & \mbox{ otherwise}.
\end{array}
\right. \label{eq:2.5i}
\end{equation}
\begin{subequations}
\begin{align}
\left[d^{\mu\nu}\right]=& \left[d_{\mu\nu}\right]^{-1}, \label{eq:2.5ii}\\
d^{\mu\nu}d_{\nu\lambda}=&\delta^{\mu}_{\;\lambda} = d_{\lambda\nu}d^{\nu\mu}. \label{eq:2.5iii}
\end{align}
\end{subequations}
For the four-dimensional vector components, the lowering of indices is accomplished by:
\begin{equation}
u_{\alpha}:=d_{\alpha\beta}u^{\beta},
\end{equation}
so that it follows:
\begin{align}
u_{i}=&d_{i\nu}u^{\nu}=\delta_{ij}u^{j}, \nonumber \\
u_{4}=&d_{4\nu}u^{\nu}=-u^{4}, \\
u^{\alpha}=&d^{\alpha\beta}u_{\beta}. \nonumber
\end{align}
The four-dimensional inner product between two vectors is provided by:
\begin{subequations}
\begin{align}
d_{\mu\nu}a^{\mu}b^{\nu}=&\delta_{ij}a^{i}b^{j}-a^{4}b^{4}, \\
d_{\mu\nu}a^{\mu}a^{\nu}=&(a^{1})^{2} +(a^{2})^{2} +(a^{3})^{2} -(a^{4})^{2}.
\end{align}
\end{subequations}
Vectors are characterized as timelike, spacelike and null by:
\begin{subequations}
\begin{align}
d_{\alpha\beta}v^{\alpha}v^{\beta} <&0 \;\;\;\; \mbox{for a timelike vector}, \\
d_{\alpha\beta}v^{\alpha}v^{\beta} >&0 \;\;\;\; \mbox{for a spacelike vector}, \\
d_{\alpha\beta}v^{\alpha}v^{\beta} =&0 \;\;\;\; \mbox{for a null
vector}.
\end{align}
\end{subequations}

\section{Particle mechanics (Newtonian and Relativistic)}
\noindent We shall start with a very brief review of Newtonian
mechanics. For the sake of simplicity, we restrict ourselves to
the case of a single point particle with mass $m > 0$. Let the
parameterized motion curve by given by
\begin{equation}
x^{i}=\mathcal{X}^{i}(t)
\end{equation}
where $x^{i}$ are Cartesian coordinates of the Euclidean space
$\mathbb{E}_{3}$ and $t$ is the time variable. Let the three
components of the force vector be given by $f^{i}(t,\mathbf{x},
\mathbf{v})$, which are functions of seven real variables. The
components $v^{i}$ represent the velocity variables.

\qquad Newton's equations of motion for a single particle are
provided by the well known equations:
\begin{equation}
m\frac{d^{2}\Chi^{i}(t)}{dt^{2}}=f^{i}(t,\mathbf{x},\,
\mathbf{v})_{|x^{i}=\Chi^{i}(t),\,v^{i}=d\Chi^{i}/dt}
\label{eq:3.2}
\end{equation}

\qquad These equations imply that
\begin{equation}
\frac{d}{dt}\left[\frac{1}{2}m \delta_{ij}\frac{d\Chi^{i}(t)}
{dt}\frac{d\Chi^{j}}{dt}\right]=
\delta_{ij}\left[v^{i}f^{j}(t,\mathbf{x},\mathbf{v})\right]_{|...}\;.
\label{eq:3.3}
\end{equation}
The above equations are physically interpreted as ``the rate of
increase of kinetic energy of the  particle is equal to the rate
of work performed by the external force''.

\qquad We note that Newton's equations of motion
(\ref{eq:3.3}) remain unchanged in form (or covariant) under
the coordinate transformations:
\begin{subequations}
\begin{align}
\hat{x}^{a}=&c^{a} +r^{a}_{\;b}x^{b}, \label{eq:3.4i}\\
\left[R\right]_{3\times 3}:=&\left[r^{a}_{\;b}\right], \label{eq:3.4ii}\\
\left[R\right]^{\mbox{\tiny{T}}}\left[R\right]=&\left[I\right]_{3\times 3}. \label{eq:3.4iii}
\end{align}
\end{subequations}
The equation (\ref{eq:3.4iii}) \emph{defines an orthogonal
matrix} $\left[R\right]_{3\times 3}$. The set of transformations
(\ref{eq:3.4i}) constitute the six parameter group
$\mathcal{IO}(3;\mathbb{R})$, the isometry group of Euclidean
three-space $\mathbb{E}_{3}$.

\qquad Consider another transformation, namely a special Galilean transformation:
\begin{align}
\hat{x}^{1}=&x^{1}-v^{1}t, \nonumber \\
\hat{x}^{2}=&x^{2}, \;\; \hat{x}^{3}=x^{3}, \label{eq:3.5} \\
\hat{t}=&t. \nonumber
\end{align}
The new (hatted) frame is moving with constant velocity $v^{1}$
along the $x^{1}$-axis relative to the old frame. The Newtonian
motion laws (\ref{eq:3.3}) remain unchanged in form by the
Galilean transformations (\ref{eq:3.5}).

\qquad In the nineteenth century, Michelson and Morley performed
some sophisticated experiments regarding light propagation in
vacuum \cite{ref:michmor}. The startling outcome of their results was that the speed
of light does \emph{not} change due to any (constant) motion of
either the source or observer. Newton's ideas of absolute space
and absolute time (inherent in (\ref{eq:3.2}) and
(\ref{eq:3.5})) are \emph{incompatible} with Michelson and
Morley's experimental findings \cite{ref:michmor}. In 1905 Einstein solved this
puzzle by the revolutionary ideas that space and time are
relative in regards to any motion \cite{ref:1}. However, a combined spacetime
continuum, $M_{4}$, is still \emph{absolute}. The appropriate
generalization of the three-dimensional Cartesian coordinates are
the four-dimensional Minkowskian coordinates \cite{ref:5} \cite{ref:6}. Moreover, the
correct generalization of the transformations in
(\ref{eq:3.4i}) are furnished by (figure 1):
\begin{subequations}
\begin{align}
\hat{x}^{\alpha}=& c^{\alpha}+l^{\alpha}_{\;\beta}x^{\beta}, \label{eq:3.6i}\\
\left[L\right]_{4\times 4}:=& \left[l^{\alpha}_{\;\beta}\right], \label{eq:3.6ii}\\
\left[D\right]_{4\times 4}\equiv& \left[d_{\mu\nu}\right], \label{eq:3.6iii} \\
\left[L\right]^{\mbox{\tiny{T}}}\left[D\right]\left[L\right]=&\left[D\right], \label{eq:3.6iv} \\
\left[d^{\mu\nu}\right]:=& \left[D\right]^{\mbox{\tiny{-1}}}
\equiv\left[D\right] =\left[d_{\mu\nu}\right]. \label{eq:3.6v}
\end{align}
\end{subequations}

\begin{figure}[ht]
\begin{center}
\includegraphics[bb=0 0 699 347, clip, scale=0.5, keepaspectratio=true]{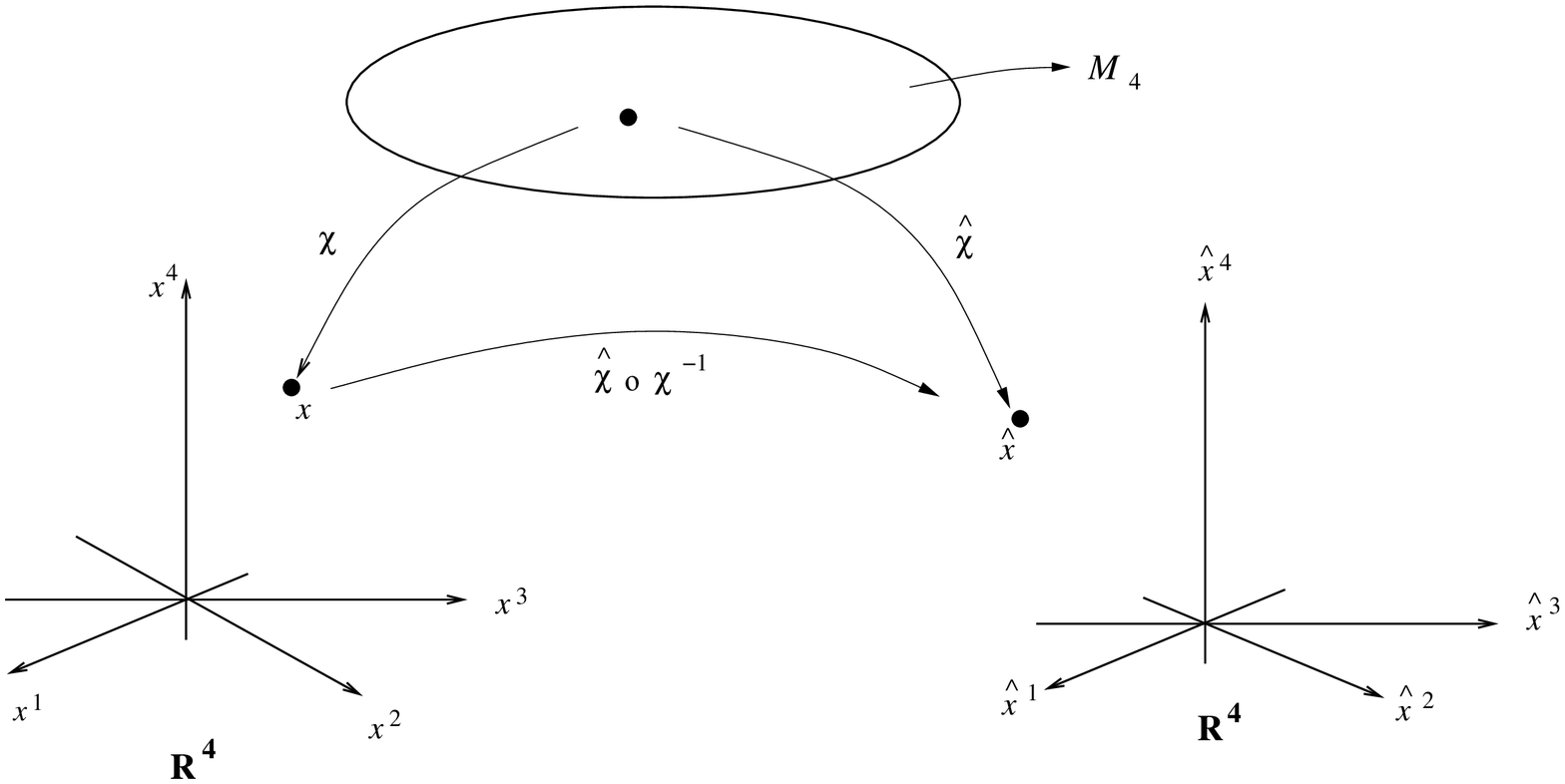}
\caption{{\small The inhomogeneous Lorentz transformation in spacetime.}} \label{fig1}
\end{center}
\end{figure}

A typical example of the above transformation (with $x^{4}=ct$) is provided by:
\begin{align}
\hat{x}^{1}=& \frac{x^{1}-\left(\frac{v^{1}}{c}\right)x^{4}}{\relgamma}=x^{1}-v^{1} t +\mathcal{O}\left(\frac{(v^{1})^{2}}{c^{2}}\right), \nonumber \\
\hat{x}^{2}=&x^{2}, \;\; \hat{x}^{3}= x^{3}, \nonumber \\
\hat{x}^{4}=&\frac{x^{4}-\left(\frac{v^{1}}{c}\right)x^{1}}{\relgamma} = ct+\mathcal{O}\left(\frac{v^{1}}{c^2}\right), \label{eq:3.7} \\
\hat{t}=& t +\mathcal{O}\left(\frac{1}{c^{2}}\right). \nonumber
\end{align}
We make the following comments on the above transformations: \\
(i) The equation (\ref{eq:3.7}) is called the Lorentz (or ``boost'') transformation \cite{ref:5} \cite{ref:6}. \\
(ii) It is the correct generalization of the Galilean transformation (\ref{eq:3.5}) for a moving frame. \\
(iii) The above transformation implies contraction of length measurements in the moving frame. \\
(iv) It can bring about the ``slowing'' of time measurements (time dilation) for a moving observer. \\
(v) The speed, $|v^{1}|$ of the moving observer must be
\emph{strictly less than} $c$, \emph{the speed of light}.

\qquad The set
of transformations in (\ref{eq:3.6i}) constitutes a continuous
group known as the inhomogeneous Lorentz group or the
Poincar\'{e} group. It is a ten parameter group denoted by
$\mathcal{IO}(3,1;\mathbb{R})$.

\qquad It follows from (\ref{eq:3.6iv}) that
$\mbox{det}\left[L\right]=\pm 1$. Therefore, the inverse matrix
exists and is denoted by
\begin{align}
\left[a^{\alpha}_{\;\beta}\right]_{4\times 4} \equiv& \left[A\right]:=\left[L\right]^{\mbox{\tiny{-1}}}, \nonumber \\
l^{\mu}_{\; \nu}a^{\nu}_{\; \beta}= & a^{\mu}_{\; \nu}l^{\nu}_{\; \beta} =\delta^{\mu}_{\; \beta}. \label{eq:3.8}
\end{align}
Now we shall define Minkowskian tensor fields in the flat
spacetime manifold $M_{4}$. These are defined by the
transformation properties \cite{ref:5}, \cite{ref:6}, \cite{ref:7}:
\begin{equation}
\hat{T}^{\alpha_{1}...
\alpha_{r}}_{\;\;\;\;\;\;\;\;\beta_{1}...\beta_{s}}(\hat{x})=
l^{\alpha_{1}}_{\;\gamma_{1}}...l^{\alpha_{r}}_{\;\gamma_{r}}
a^{\mu_{1}}_{\;\beta_{1}}...a^{\mu_{s}}_{\;\beta_{s}}
T^{\gamma_{1}...\gamma_{r}}_{\;\;\;\;\;\;\;\;\mu_{1}...\mu_{s}}(x).
\label{eq:3.9}
\end{equation}
Here, the coefficients $l^{\alpha}_{\;\gamma}$,
$a^{\mu}_{\;\beta}$ are defined by (\ref{eq:3.6i}) and
(\ref{eq:3.8}). The tensor fields in (\ref{eq:3.9})
are of order $r+s$, where $r$ is the contravariant order and $s$
is the covariant order. Note that the right hand side of
(\ref{eq:3.9}) condenses a sum of $4^{r+s}$ terms!
Moreover, these tensor fields are assumed to be twice
continuously differentiable. The restriction of the tensor fields
on a parameterized curve in $M_{4}$ satisfies the transformation
rules:
\begin{equation}
\hat{T}^{\alpha_{1}...\alpha_{r}}_{\;\;\;\;\;\;\;\;\beta_{1}...\beta_{s}}
(\hat{x})_{|\hat{x}=\hat{\Chi}(\tau)}=
l^{\alpha_{1}}_{\;\gamma_{1}}...l^{\alpha_{r}}_{\;\gamma_{r}}
a^{\mu_{1}}_{\;\beta_{1}}...a^{\mu_{s}}_{\;\beta_{s}}
T^{\gamma_{1}...\gamma_{r}}_{\;\;\;\;\;\;\;\;\mu_{1}...\mu_{s}}(x)_{|{x}={\Chi}(\tau)}.
\label{eq:3.10}
\end{equation}
As simple examples, we consider the numerical second order
tensors $d_{\mu\nu}$, $d^{\alpha\beta}$ in (\ref{eq:3.6iii}) and
(\ref{eq:3.6v}) respectively. By the rules
(\ref{eq:3.9}) we deduce that
\begin{align}
\hat{d}_{\alpha\beta}=&a^{\mu}_{\;\alpha}a^{\nu}_{\;\beta}d_{\mu\nu}
=d_{\alpha\beta}, \nonumber \\
\hat{d}^{\alpha\beta}=&l^{\alpha}_{\;\mu}l^{\beta}_{\;\nu}d^{\mu\nu}
=d^{\alpha\beta}. \label{eq:3.11}
\end{align}
These special tensor components \emph{retain} their numerical
values under the rules (\ref{eq:3.10}).

\qquad Now, let us consider a $0+0$ order or scalar field $W(x)$
which is twice differentiable. Further, let it satisfy the wave
equation:
\begin{equation}
\square W(x):=d^{\alpha\beta}\frac{\partial^{2}}{\partial
x^{\alpha}\partial
x^{\beta}}=\left[\frac{\partial^{2}}{\left(\partial
x^{1}\right)^{2}}+ \frac{\partial^{2}}{\left(\partial
x^{2}\right)^{2}} + \frac{\partial^{2}}{\left(\partial
x^{3}\right)^{2}} - \frac{1}{c^{2}}
\frac{\partial^{2}}{\left(\partial t\right)^{2}} \right]W(..)=0.
\label{eq:3.12}
\end{equation}
We can prove from (\ref{eq:3.10}) (which in this case
reads $\hat{W}(\hat{x})=W(x)$), (\ref{eq:3.11}), (\ref{eq:3.6i})
and the chain rule of differentiation that (\ref{eq:3.12})
implies:
\begin{equation}
\hat{\square}\hat{W}(\hat{x})=\hat{d}^{\alpha\beta}
\frac{\partial^{2}\hat{W}(\hat{x})}{\partial\hat{x}^{\alpha}
\partial \hat{x}^{\beta}} =  \left[\frac{\partial^{2}}{\left(\partial
\hat{x}^{1}\right)^{2}}+ \frac{\partial^{2}}{\left(\partial
\hat{x}^{2}\right)^{2}} + \frac{\partial^{2}}{\left(\partial
\hat{x}^{3}\right)^{2}} - \frac{1}{c^{2}}
\frac{\partial^{2}}{\left(\partial \hat{t}\right)^{2}}
\right]\hat{W}(..) = 0. \label{eq:3.13}
\end{equation}
Since the speeds of wave propagation in (\ref{eq:3.12}) and
(\ref{eq:3.13}) are both $c$, the speed of light, we conclude
that the speed of such a wave remains invariant under any motion
of an observer characterized by (\ref{eq:3.7}).

\qquad Now we discuss another important feature of the tensor
field in (\ref{eq:3.9}) (or (\ref{eq:3.10})). The
four-dimensional tensor field equation
\begin{equation}
T^{\gamma_{1}...\gamma_{r}}_{\;\;\;\;\;\;\;\;
\mu_{1}...\mu_{s}}(x)= 0
\end{equation}
hold if and only if the transformed components satisfy
\begin{equation}
\hat{T}^{\alpha_{1}...\alpha_{r}}_{\;\;\;\;\;\;\;\;
\beta_{1}...\beta_{s}}(\hat{x}) =0.
\end{equation}
This statement physically signifies that a natural law
expressible by the vanishing of a tensor field remains unaltered in any
rotated, reflected or moving frame. The main mathematical
postulate of the special theory of relativity is that the natural
laws must be expressed as tensor field equations in spacetime.

\qquad Now we shall study the relativistic particle mechanics.
For that purpose we have to introduce the exact definition of a
parameterized curve in $M_{4}$ and its physical interpretation. It
is easier to consider the corresponding parameterized curve in the
coordinate space $\mathbb{R}^{4}$. See figure 2.

\begin{figure}[ht]
\begin{center}
\includegraphics[bb=50 323 618 608, clip, scale=0.5, keepaspectratio=true]{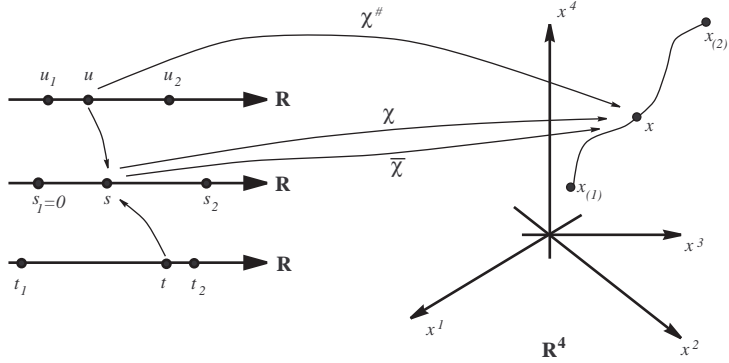}
\caption{{\small A parametrized curve in spacetime.}} \label{fig2}
\end{center}
\end{figure}

\qquad Let a differentiable parameterized curve into the
Minkowskian coordinate space $\mathbb{R}^{4}$ be characterized be
characterized by
\begin{align}
x^{\mu}=&\Chi^{\#\, \mu}(u) , \label{eq:3.14} \\
u_{1} \leq& u \leq u_{2}\, . \nonumber
\end{align}
Here, the functions $\Chi^{\#\,\mu}$ are assumed to be
continuously twice differentiable. Suppose that the curve
represents physically the history of an idealized ``point'' clock
moving in spacetime. The \emph{proper} (or actual) time flow of
the clock along the motion curve is given by \cite{ref:5},
\cite{ref:6}
\begin{equation}
s=S^{\#}(u):=\frac{1}{c} \int_{u_{1}}^{u} \sqrt{-d_{\alpha\beta}
\frac{d\Chi^{\#\,\alpha}(w)}{dw}
\frac{d\Chi^{\#\,\beta}(w)}{dw}}\,dw. \label{eq:3.15}
\end{equation}
Here, we have tacitly assumed that
\begin{equation}
 0 < \sqrt{-d_{\alpha\beta}
\frac{d\Chi^{\#\,\alpha}(u)}{du} \frac{d\Chi^{\#\,\beta}(u)}{du}}
\equiv c \frac{dS^{\#}(u)}{du} \label{eq:3.16}
\end{equation}
for all $u \in [u_{1},\;u_{2}]$. The suppositions above are based
on physical principles. Consider for example a free particle in
the spacetime. The history curve, or world line, of such a
particle would be a straight line with some slope. An observer
``traveling'' with this particle, feeling no acceleration can
justifiably claim that he or she is not moving but it is the
surroundings which are moving. Therefore, to the observer, he or
she is not moving in space but only in time. Generalizing this, a
straight time-like world line therefore represents the appropriate time
axis for an observer on this trajectory.

\qquad Physically, (\ref{eq:3.16}) implies that the curve is
timelike and the actual speed along the curve, as measured by an
observer whose time axis is given by the $x^{4}$ axis in figure
2, is \emph{always less than the speed of light}.

\qquad The integral (\ref{eq:3.15}), which defines the proper
time, $s$, is \emph{invariant} (or scalar) with respect to
transformations (\ref{eq:3.6i}). Moreover, the integral
(\ref{eq:3.15}) is \emph{invariant} under any smooth
reparameterization of the curve characterized by
\begin{align}
x^{\mu}=&\Chi^{\#\,\mu}(u)=\tilde{\Chi}^{\mu}(y), \nonumber \\
y=&Y(u), \; \frac{dY(u)}{du} \neq 0, \nonumber \\
Y(u_{1}):=&y_{1} \leq y \leq y_{2} := Y(u_{2}). \label{eq:3.17}
\end{align}
Chosing the parameter $y=s$, the proper time, we obtain from
(\ref{eq:3.15}), (\ref{eq:3.16}) and (\ref{eq:3.17}) that
\begin{subequations}
\begin{align}
x^{\mu}=&\Chi^{\#\,\mu}(u)=:\Chi^{\mu}(s), \;\; s_{1}=0 \leq s
\leq s_{2}, \label{eq:3.18i} \\
s=& S(s) =\frac{1}{c} \int_{0}^{s}\sqrt{-d_{\alpha\beta}
\frac{d\Chi^{\alpha}(w)}{dw} \frac{d\Chi^{\beta}(w)}{dw}}\,dw,
\label{eq:3.18ii} \\
1\equiv& \frac{dS(s)}{ds} =\frac{1}{c} \sqrt{-d_{\alpha\beta}
\frac{d\Chi^{\alpha}(s)}{ds} \frac{d\Chi^{\beta}(s)}{ds}},
\label{eq:3.18iii} \\
&d_{\alpha\beta} \frac{d\Chi^{\alpha}(s)}{ds}
\frac{d\Chi^{\beta}(s)}{ds} \equiv -c^{2} < 0 . \label{eq:3.18iv}
\end{align}
\end{subequations}
Here, $\frac{d\Chi^{\alpha}(s)}{ds}$ are the four components of
the \emph{relativistic} velocity along the motion curve.

\qquad In case the motion curve $x^{\mu}=\Chi^{\mu}(s)$ is
continuously twice differentiable, which we shall always assume,
the differentiation of (\ref{eq:3.18iv}) yields
\begin{equation}
d_{\alpha\beta}\frac{d\Chi^{\alpha}(s)}{ds}
\frac{d^{2}\Chi^{\beta}(s)}{ds^{2}} \equiv 0. \label{eq:3.19}
\end{equation}
Therefore, in the proper time parametrization, \emph{the
four-acceleration is always (Minkowskian) orthogonal to the
four-velocity!}

\qquad In case we reparameterize the curve by choosing $y=t$, the
usual or coordinate time, we derive from (\ref{eq:3.15}),
(\ref{eq:3.16}) and (\ref{eq:3.17}) that
\begin{subequations}
\begin{align}
x^{\mu}=\Chi^{\mu\#}(u)=:&\overline{\Chi}^{\mu}(t),\;
x^{i}=\overline{\Chi}^{i}(t),\; x^{4}=\overline{\Chi}^{4}(t):=ct,
\label{eq:3.20.i}
\\
s=&\overline{S}(t) =\frac{1}{c}\int_{t_{1}}^{t}
\sqrt{-d_{\alpha\beta} \frac{d\overline{\Chi}^{\alpha}(w)}{dw}
\frac{d\overline{\Chi}^{\beta}(w)}{dw}}\,dw, \label{eq:3.20ii} \\
\frac{d\overline{S}(t)}{dt} \equiv& \frac{1}{c} \sqrt{-d_{\alpha\beta}
\frac{d\overline{\Chi}^{\alpha}(t)}{dt}
\frac{d\overline{\Chi}^{\beta}(t)}{dt}} > 0, \label{eq:3.20iii} \\
\left[\frac{d\overline{S}(t)}{dt}\right]^{2} \equiv &1 -
\frac{1}{c^{2}}\left[\delta_{ij}
\frac{d\overline{\Chi}^{i}(t)}{dt}
\frac{d\overline{\Chi}^{j}(t)}{dt}\right]. \label{eq:3.20iv}
\end{align}
\end{subequations}
Recall that the Newtonian 3-velocity variables in
(\ref{eq:3.2}) are
\begin{align}
v^{i}=&V^{i}(t):=\frac{d\overline{\Chi}^{i}(t)}{dt}, \nonumber \\
|\mathbf{v}|^{2}=&|\mathbf{V}(t)|^{2}=\delta_{ij} V^{i}(t)
V^{j}(t) \geq 0. \label{eq:3.21}
\end{align}
Thus, we deduce from (\ref{eq:3.20iv}) and (\ref{eq:3.21}) that
\begin{align}
\left[\frac{d\overline{S}(t)}{dt}\right]^{2}=&
1-\frac{|\mathbf{V}(t)|^{2}}{c^{2}}, \nonumber \\
0 < \frac{d\overline{S}(t)}{dt}=& +
\sqrt{1-\frac{|\mathbf{V}(t)|^{2}}{c^{2}}} \leq 1. \label{eq:3.22}
\end{align}

\qquad Now we investigate the relationships among Newtonian
3-velocity components with the correcponding relativistic
4-velocity components. Using the chain rule of differentiation,
we obtain from (\ref{eq:3.18i}, \ref{eq:3.18ii},
\ref{eq:3.18iii}), (\ref{eq:3.20.i}, \ref{eq:3.20ii},
\ref{eq:3.20iii}), (\ref{eq:3.21}) and (\ref{eq:3.22}) that
\begin{subequations}
\begin{align}
\frac{d\overline{\Chi}^{i}(t)}{dt}=& \frac{d\overline{S}(t)}{dt}
\frac{d{\Chi}^{i}(s)}{ds}, \label{eq:3.23i} \\
\frac{d\Chi^{i}(s)}{ds}=& \frac{V^{i}(t)}{\genrelgamma},
\label{eq:3.23ii} \\
\frac{d\Chi^{4}(s)}{ds}=& \frac{1}
{\left[\frac{d\overline{S}(t)}{dt}\right]}
\frac{d\overline{\Chi}^{4}(t)}{dt} =\frac{c}{\genrelgamma} > c.
\label{eq:3.23iii}
\end{align}
\end{subequations}

\qquad Now we shall generalize the Newtonian equations of motion
(\ref{eq:3.2}) into the relativistic arena. We postulate, as
a generalization of (\ref{eq:3.2}), the special relativistic
equations of motion as \cite{ref:5}, \cite{ref:6}:
\begin{equation}
m\frac{d^{2}\Chi^{\alpha}(s)}{ds^{2}}=
\mathcal{F}^{\alpha}(x;\,u)_{|x^{\alpha}=\Chi^{\alpha}(s),\;u^{\alpha}=\frac{d\Chi^{\alpha}(s)}{ds}}.
\label{eq:3.24}
\end{equation}
Here, $(x):=(x^{1}, \, x^{2},\,x^{3}, \, x^{4})$ and
$(u):=(u^{1},\, u^{2}, \, u^{3}, \, u^{4})$. Each of the four
components $\mathcal{F}^{\alpha}$ is a function of eight
real variables and represents physically the relativistic force.
(We have tacitly assumed that the mass , $m >0$, remains
\emph{unchanged} along the motion curve in spacetime.) It follows
from (\ref{eq:3.19}), (\ref{eq:3.23ii}, \ref{eq:3.23iii}) and
(\ref{eq:3.24}) that
\begin{subequations}
\begin{align}
&d_{\alpha\beta}\mathcal{F}^{\alpha}(..)_{|..}
\frac{d\Chi^{\beta}(s)}{ds} \equiv 0, \label{eq:3.25i} \\
\mathcal{F}^{4}(..)_{|..} \equiv& \delta_{ij} \mathcal{F}^{i}(..)_{|..}
\frac{d\Chi^{j}(s)}{ds} \left[\frac{d\Chi^{4}(s)}{ds}\right]^{-1}
\equiv \delta_{ij}F^{i}(..)\frac{V^{j}(t)}{c} . \label{eq:3.25ii}
\end{align}
\end{subequations}
Now we shall compare the relativistic equations (\ref{eq:3.24})
with the Newtonian equations (\ref{eq:3.2}) and
(\ref{eq:3.3}). Let us digress slightly. Suppose that we have
a differentiable function $f$ defined along the motion curve. By
the equations (\ref{eq:3.18i}), (\ref{eq:3.20.i}) and
(\ref{eq:3.22}) and the chain rule, we get
\begin{subequations}
\begin{align}
\overline{f}(t):=& f\left[\overline{S}(t)\right]=f(s) , \label{eq:3.26i} \\
\frac{df(s)}{ds}=& \frac{1}{\genrelgamma}
\frac{d\overline{f}(t)}{dt} \label{eq:3.26ii}
\end{align}
\end{subequations}
Substituting (\ref{eq:3.26ii}) into the first three equations in
(\ref{eq:3.24}) we deduce that
\begin{subequations}
\begin{align}
\frac{d}{dt} \left[\frac{m V^{i}(t)}{\genrelgamma} \right] =&
\genrelgamma \mathcal{F}^{i}(..)_{..}, \label{eq:3.27i} \\
m\frac{d^{2}\overline{\Chi}^{i}(t)}{dt^{2}} =& \mathcal{F}^{i}
(..)_{|..} + \mathcal{O}\left(\frac{1}{c^{2}}\right).
\label{eq:3.27ii}
\end{align}
\end{subequations}
Comparing (\ref{eq:3.27i}) with (\ref{eq:3.2}) we conclude
that the Newtonian momentum components, $mV^{i}(t)$ have to be
modified into $mV^{i}(t)/\genrelgamma$ in relativity. Moreover,
the Newtonian force components relate to relativistic force
components by the equations:
\begin{equation}
f^{i}(t,\mathbf{x},\mathbf{v})_{x^{i}=\overline{\Chi}^{i}(t),v^{i}=
\frac{d\overline{\Chi}^{i}(t)}{dt}} = \genrelgamma
\mathcal{F}^{i}(x, u)_{x^{i}=\overline{\Chi}^{i}(t), x^{4}=ct,
u^{i}=v^{i}/{\sqrt{..}}, u^{4} =c/{\sqrt{..}}} \label{eq:3.28}
\end{equation}
The fourth equation in (\ref{eq:3.24}) yields, with
(\ref{eq:3.23iii}), (\ref{eq:3.25ii}), (\ref{eq:3.26ii}) and
(\ref{eq:3.28})
\begin{subequations}
\begin{align}
m\frac{d}{dt} \left[\frac{c}{\genrelgamma} \right]=& \genrelgamma
\mathcal{F}^{4}(..)_{|..}, \label{eq:3.29i} \\
\mbox{or, }\;\;\; \frac{d}{dt} \left[\frac{mc^{2}}{\genrelgamma}
\right]=& \delta_{ij} \left[v^{i}f^{j}(..)\right]_{|..}\; .
\label{eq:3.29ii}
\end{align}
\end{subequations}
Expanding (\ref{eq:3.29ii}) for $|\mathbf{V}(t)|/c < 1$, we obtain
that
\begin{equation}
\frac{d}{dt} \left[mc^{2} +\frac{1}{2} m |V(t)|^{2} + \mathcal{O}
\left(\frac{1}{c^{2}}\right) \right] = \delta_{ij} \left[v^{i}
f^{j}(..)\right]_{|..} . \label{eq:3.30}
\end{equation}
Comparing the above equation with the corresponding Newtonian
equation (\ref{eq:3.3}), we conclude that the instantaneous
energy $E(|\mathbf{V}(t)|)$ of the particle must be furnished by
\begin{equation}
E(|\mathbf{V}(t)|)=\frac{mc^{2}}{\genrelgamma} =mc^{2}
+\frac{m}{2} |\mathbf{V}(t)|^{2}
+\mathcal{O}\left(\frac{1}{c^{2}}\right) . \label{eq:3.31}
\end{equation}
In the limit $|\mathbf{V}| \rightarrow 0$, we derive that
\begin{equation}
E(0)=mc^{2}\, . \label{eq:3.32}
\end{equation}
The above equation, which is the most famous formula of modern
science, reveals the \emph{enormous rest energy} of a massive
particle. (The rest energy associated with a single $1$ kilogram object
is approximately $9\times 10^{16}$ Joules, enough to meet New York City's energy requirements for more than seven months!)

\section{Electromagnetic fields}
Maxwell's equations of electromagnetic fields are, in Gaussian units, the following:
\begin{subequations}
\begin{align}
c\:\epsilon_{ijk} \frac{\partial B^{j}(\mathbf{x},t)}{\partial
x^{k}}=& -4\pi j_{i}(\mathbf{x},t) -
\frac{\partial E_{i}(\mathbf{x}, t)}{\partial
t}, \label{eq:4.1i} \\
\frac{\partial E^{i}(\mathbf{x},t)}{\partial x^{i}}=& 4\pi\sigma
(\mathbf{x},t),
\label{eq:4.1ii} \\
\frac{\partial B^{i}(\mathbf{x},t)}{\partial x^{i}} =&0,
\label{eq:4.1iii} \\
c\:\epsilon_{ijk} \frac{\partial E^{j}(\mathbf{x},t)}{\partial
x^{k}}=& \frac{\partial{B}^{i}(\mathbf{x},t)}{\partial t},
\label{eq:4.1iv}
\end{align}
\end{subequations}
Here, $E^{i}(..)$ and $B^{i}(..)$ stand for the electric and
magnetic field components respectively. Also,  The charge density
and the current density components are denoted by $\sigma(..)$
and $j_{i}(..)$ respectively.

\qquad Another popular system of units, which we shall \emph{not}
employ here, is the Systeme Internationale (SI). In these units,
Maxwell's equations read:
\begin{subequations}
\begin{align}
\epsilon_{ijk} \frac{\partial B^{j}(\mathbf{x},t)}{\partial
x^{k}}=& -\mu_{0} j_{i}(\mathbf{x},t) -\frac{1}{c^{2}}
\frac{\partial E_{i}(\mathbf{x}, t)}{\partial
t}, \label{eq:4.1SIi} \\
\frac{\partial E^{i}(\mathbf{x},t)}{\partial x^{i}}=&
\frac{\sigma (\mathbf{x},t)}{\epsilon_{0}},
\label{eq:4.1SIii} \\
\frac{\partial B^{i}(\mathbf{x},t)}{\partial x^{i}} =&0,
\label{eq:4.1SIiii} \\
\epsilon_{ijk} \frac{\partial E^{j}(\mathbf{x},t)}{\partial
x^{k}}=& \frac{\partial{B}^{i}(\mathbf{x},t)}{\partial t}.
\label{eq:4.1SIiv}
\end{align}
\end{subequations}
The constants
$\epsilon_{0}$ and $\mu_{0}$ respectively represent the
permittivity and permeability of free space with the relation $c=1/\sqrt{\epsilon_{0}\mu_{0}}$.

\qquad The equations (\ref{eq:4.1i} -
\ref{eq:4.1iv}) imply that the charge current conservation
equation:
\begin{equation}
\frac{\partial j^{i}(\mathbf{x},t)}{\partial x^{i}} +
\frac{\partial \sigma(\mathbf{x},t)}{\partial t} =0 \label{eq:4.2}
\end{equation}
must be satisfied.

\qquad The energy density of the electromagnetic field is
characterized by
\begin{equation}
u(\mathbf{x},t):= \frac{1}{8\pi} \delta_{ij}
\left[E^{i}(..)E^{j}(..) +  B^{i}(..)B^{j}(..) \right]
. \label{eq:4.3}
\end{equation}
The momentum density of the electromagnetic field is provided by
\begin{equation}
S_{i}(\mathbf{x},t):=\frac{1}{4\pi} \epsilon_{ijk}E^{j}(\mathbf{x},t)
B^{k}(\mathbf{x},t) \label{eq:4.4}
\end{equation}
The corresponding vector field $\mathbf{S}(\mathbf{x},t)$ is also known as the
\emph{Poynting vector}.

\qquad Maxwell's \emph{electromagnetic stress tensor} is
furnished by
\begin{align}
M_{ij}(\mathbf{x},t)=&
\frac{1}{4\pi} \left\{\left[E_{i}(\mathbf{x},t)E_{j}(\mathbf{x},t)
-\frac{1}{2}\delta_{ij}
\delta^{kl}E_{k}(\mathbf{x},t)E_{l}(\mathbf{x},t)\right] \right.
\nonumber \\
&+ \left.\left[B_{i}(\mathbf{x},t)B_{j}(\mathbf{x},t)
+\frac{1}{2}\delta_{ij}
\delta^{kl}B_{k}(\mathbf{x},t)B_{l}(\mathbf{x},t)\right] \right\}
\label{eq:4.5}
\end{align}

\qquad The components of the \emph{Lorentz force} on a charged
particle (of net charge $e$) is given by
\begin{equation}
f_{i}(t,\mathbf{x}, \mathbf{v}):=e \left[E_{i}(\mathbf{x},t) +
\epsilon_{ijk}\frac{v^{j}}{c}B^{k}(\mathbf{x},t)\right]. \label{eq:4.6}
\end{equation}
Note that this equation yields
\begin{equation}
v^{i}f_{i}(t,\mathbf{x},\mathbf{v})=eE_{i}(\mathbf{x},t)v^{i} ,
\label{eq:4.7}
\end{equation}
indicating the well known result that the magnetic field makes
no contribution to the rate of work.

\qquad Now, we shall obtain the special relativistic versions of
the various electromagnetic equations. Following Minkowski
\cite{ref:9}, we define the four-dimensional electomagnetic
field tensor components as:
\begin{equation}
\left[F_{\mu\nu}(x)\right]:= \left[
\begin{array}{cccc}
0 & B_{3}(x)& -B_{2}(x)& E_{1}(x) \\
-B_{3}(x) & 0 & B_{1}(x) & E_{2}(x) \\
B_{2}(x)& -B_{1}(x) & 0 & E_{3}(x) \\
-E_{1}(x) & -E_{2}(x) & -E_{3}(x) & 0
\end{array}
\right] \equiv\left[-F_{\nu\mu}(x)\right]. \label{eq:4.8}
\end{equation}
It should be mentioned that Minkowski first \emph{unified}
electric and magnetic fields by the definition (\ref{eq:4.8}) of
the four-dimensional anti-symmetric tensor $F_{\mu\nu}(x)$. The
above is one of several possible definitions for the tensor
$F_{\mu\nu}$. We define the relativistic charge-current components
by:
\begin{align}
J^{i}(x):=&j^{i}(\mathbf{x},t), \nonumber \\
J^{4}(x):=&c\sigma(\mathbf{x},t). \label{eq:4.9}
\end{align}
The Maxwell equations (\ref{eq:4.1i}-\ref{eq:4.1iv}), along with
definitions (\ref{eq:4.8}) and (\ref{eq:4.9}), neatly boil down to
\cite{ref:8}
\begin{subequations}
\begin{align}
\frac{\partial F^{\mu\nu}(x)}{\partial x^{\nu}}&= \frac{4\pi}{c}J^{\mu}(x),
\label{eq:4.10i} \\
\frac{\partial F^{\mu\nu}(x)}{\partial x^{\lambda}}+\frac{\partial
F^{\nu\lambda}(x)}{\partial x^{\mu}}+\frac{\partial
F^{\lambda\mu}(x)}{\partial x^{\nu}}&=0 \label{eq:4.10ii} \\
\frac{\partial J^{\mu}(x)}{\partial x^{\mu}}&=0 .
\label{eq:4.10iii}
\end{align}
\end{subequations}
Outside of charged matter, Maxwell's equations are summarized by:
\begin{align}
\frac{\partial F^{\mu\nu}(x)}{\partial x^{\nu}}=& 0, \nonumber \\
\frac{\partial F_{\mu\nu}(x)}{\partial x^{\lambda}}+\frac{\partial
F_{\nu\lambda}(x)}{\partial x^{\mu}}+\frac{\partial
F_{\lambda\mu}(x)}{\partial x^{\nu}}&=0, \label{eq:4.11}
\end{align}
which together imply
\begin{equation}
\square F_{\mu\nu}(x)= \left[\frac{\partial^{2}}{\left(\partial
x^{1}\right)^{2}}+ \frac{\partial^{2}}{\left(\partial
x^{2}\right)^{2}}+ \frac{\partial^{2}}{\left(\partial
x^{3}\right)^{2}}- \frac{1}{c^{2}}
\frac{\partial^{2}}{\left(\partial t\right)^{2}} \right] F_{\mu\nu}(x) =0.
\label{eq:4.11b}
\end{equation}
Since the above equations are tensor field equations in the four
dimensional spacetime, Maxwell's equations outside matter were
already relativistic even before Einstein's discovery of the
special theory of relativity! Moreover, the equation
(\ref{eq:4.11b}) implies that electromagnetic waves propagate
with the speed of light. Since the wave operator $\square$ is a
relativistic invariant (see equation (\ref{eq:3.13})), we can
conclude that the speed of electromagnetic wave propagation
remains unchanged under the boost transformation (\ref{eq:3.6i})
characterizing a moving observer. Thus, Michelson and Morley's
experimental puzzle is logically explained.

\qquad Now we shall \emph{unify} energy density, momentum density
and Maxwell's stress tensor by defining a relativistic
electromagnetic energy-momentum-stress tensor:
\begin{subequations}
\begin{align}
\mathcal{M}^{\alpha\beta}(x):=&\frac{1}{4\pi} \left[F^{\lambda\alpha}(x)
F_{\lambda}^{\;\;\beta}(x) -\frac{1}{4}d^{\alpha\beta}F_{\mu\nu}(x)F^{\mu\nu}(x)\right]
\equiv \mathcal{M}^{\beta\alpha} , \label{eq:4.12i} \\
\left[\mathcal{M}^{\alpha\beta}(x)\right] =& \left[
\begin{array}{cc}
-M^{ij}(\mathbf{x},t) & cS_{i}(\mathbf{x},t) \\
cS_{i}(\mathbf{x},t) & u(\mathbf{x},t)
\end{array}
\right]. \label{eq:4.12ii}
\end{align}
\end{subequations}
Working out the covariant divergence, $\frac{\partial
M_{\alpha}^{\;\;\beta}}{\partial x^{\beta}}$, using
(\ref{eq:4.10i}), (\ref{eq:4.10ii}) and (\ref{eq:4.12i}) yields
\begin{align}
\frac{\partial \mathcal{M}_{\alpha}^{\;\;\beta}}{\partial x^{\beta}} = & \frac{1}{4\pi}\left[F^{\lambda}_{\;\;\alpha} \frac{\partial F_{\lambda}^{\;\;\beta}}{\partial x^{\beta}} + F^{\;\;\beta}_{\lambda} \frac{\partial F_{\;\;\alpha}^{\lambda}}{\partial x^{\beta}} - \frac{1}{2} F^{\mu\nu} \frac{\partial F_{\mu\nu}}{\partial x^{\alpha}} \right]\nonumber \\
=&\frac{1}{c}F^{\lambda}_{\;\; \alpha} J_{\lambda} +\frac{1}{8\pi} F^{\lambda\beta} \left[\frac{\partial F_{\lambda\alpha}}{\partial x^{\beta}} + \frac{\partial F_{\alpha\beta}}{\partial x^{\lambda}} + \frac{\partial F_{\beta\lambda}}{\partial x^{\alpha}}\right] \label{eq:4.13} \\
=& \frac{1}{c}F^{\lambda}_{\;\;\alpha} J_{\lambda} + 0. \nonumber
\end{align}

\qquad Now, the relativistic Lorentz equation for a charged
particle with mass $m$  and charge $e$ are taken to be (compare
with (\ref{eq:3.24}))
\begin{subequations}
\begin{align}
m\frac{d^{2}\Chi^{\alpha}(s)}{ds^{2}}=&\mathcal{F}^{\alpha}(..)_{|..}:= \frac{e}{c}F^{\alpha}_{\;\lambda}(x) _{|\Chi(s)} \frac{d\Chi^{\lambda}(s)}{ds}, \label{eq:4.14i} \\
\mathcal{F}_{\alpha}(..)_{|..} \frac{d\Chi^{\alpha}(s)}{ds}=& \frac{e}{c}F_{\alpha \lambda}(x)_{|..} \frac{d\Chi^{\alpha}(s)}{ds} \frac{d\Chi^{\lambda}(s)}{ds} \equiv 0, \label{eq:4.14ii} \\
\genrelgamma \mathcal{F}_{i}(..)_{|..} =& \frac{e}{c}\left[ F_{ij}(\mathbf{x},t)_{|\overline{\Chi}(t)} \frac{d\overline{\Chi}^{j}(t)}{dt} + c F_{14}(..)_{|..}\right] \nonumber \\
 =&e \left[E_{i}(\mathbf{x},t) + \epsilon_{ijk}\frac{v^{j}}{c}(t)B^{k}(\mathbf{x},t)\right]_{|\overline{\Chi}(t)}. \label{eq:4.14iii}
\end{align}
\end{subequations}
The right hand side of (\ref{eq:4.14iii}) yields the correct
Lorentz force given in equation (\ref{eq:4.6}).

\qquad Now we shall briefly introduce the electromagnetic four-potential. According to the converse Poincar\'{e} lemma
\cite{ref:10}, the equations (\ref{eq:4.10ii}) imply that
there exists relativistic field components $A_{\mu}(x)$ of class
$C^{3}$ such that
\begin{equation}
F_{\mu\nu}(x)= \frac{\partial}{\partial{x^{\mu}}}A_{\nu}(x)
-\frac{\partial}{\partial x^{\nu}} A_{\mu}(x), \label{eq:4.15}
\end{equation}
so that
\begin{subequations}
\begin{align}
E_{i}(\mathbf{x},t)=&\frac{\partial A_{4}}{\partial x^{i}} -\frac{1}{c} \frac{\partial A_{i}}{\partial t}, \label{eq:4.15ii} \\
B^{i}(\mathbf{x},t)=& \frac{1}{2}\epsilon^{ijk} \left[\frac{\partial A_{k}}{\partial x^{j}} - \frac{\partial A_{j}}{\partial x^{k}} \right]. \label{eq:4.15iii}
\end{align}
\end{subequations}
We note that the four-potential components $A_{\mu}(x)$ are not
unique. We can make a (local) gauge transformation:
\begin{align}
A^{\prime}_{\mu}(x)=& A_{\mu}(x) -\frac{\partial \Lambda(x)}{\partial x^{\mu}}, \label{eq:4.16} \\
F^{\prime}_{\mu\nu}(x) \equiv &F_{\mu\nu}(x) . \nonumber
\end{align}
Here, $\Lambda(x)$ is an \emph{arbitrary} function of class
$C^{4}$. We can perform a gauge transformation such that the
function $\Lambda(x)$ satisfies the partial differential equation
\begin{equation}
\square \Lambda(x)=\frac{\partial A^{\mu}}{\partial x^{\mu}} . \label{eq:4.17}
\end{equation}
Thus, from equation (\ref{eq:4.16}) and (\ref{eq:4.17}), we get
\begin{equation}
\frac{\partial A^{\prime\,\mu}}{\partial x^{\mu}}= \frac{\partial
A^{\mu}}{\partial x^{\mu}} - \square \Lambda =0. \label{eq:4.18}
\end{equation}
The above condition on $A^{\prime\,\mu}$ is called the
\emph{Lorentz gauge condition}. In this gauge, Maxwell's
equations (\ref{eq:4.10i} - \ref{eq:4.10iii}) reduce, by
(\ref{eq:4.16}) to
\begin{subequations}
\begin{align}
d^{\mu\lambda} \frac{\partial}{\partial x^{\lambda}}
\left[\frac{\partial A^{\prime \, \nu}}{\partial x^{\nu}} \right]
-\square A^{\prime \, \mu} =& -\square A^{\prime \, \mu}= 4\pi J^{\mu}(x) \label{eq:4.19i} \\
\frac{\partial J^{\mu}(x)}{\partial x^{\mu}} =&0.
\label{eq:4.19ii}
\end{align}
\end{subequations}
The relativistic \emph{inertial} energy-momentum-stress tensor for
the charged matter is given by:
\begin{subequations}
\begin{align}
\left[\mathcal{I}^{\mu\nu}(x)\right]:= \left[
\begin{array}{cc}
\rho(x)U^{i}(x)U^{j}(x) & \rho(x) U^{i}(x)U^{4}(x) \\
\rho(x) U^{i}(x)U^{4}(x) & \rho(x) \left(U^{4}(x)\right)^{2}
\end{array}
\right] \equiv & \left[ \mathcal{I}^{\nu\mu}(x) \right],
\label{eq:4.20i} \\
d_{\mu\nu} U^{\mu}(x) U^{\nu}(x) \equiv& -c^{2} .
\label{eq:4.20ii}
\end{align}
\end{subequations}
Here, $\rho(x)$ represents the invariant (or proper) mass density
and $U^{\mu}(x)$ are components of the four-velocity field. Thus,
$\rho(x) U^{i}(x)$ are related to the three-momentum density
components and $\rho(x)\left[U^{4}(x)\right]^{2}$ is the energy density.

\qquad Observing the similarity between $\mathcal{M}^{\alpha\beta}(x)$ in
(\ref{eq:4.12ii}) and $\mathcal{I}^{\alpha\beta}(x)$ in
(\ref{eq:4.20i}), we define the \emph{total energy-momentum
stress field} for a charged material as
\begin{equation}
T^{\alpha\beta}(x):=\mathcal{I}^{\alpha\beta}(x) +
\mathcal{M}^{\alpha\beta}(x) \equiv \mathcal{I}^{\beta\alpha}(x) +
\mathcal{M}^{\beta\alpha}(x)\equiv T^{\beta\alpha}(x),
\label{eq:4.21}
\end{equation}
Using (\ref{eq:4.12i}), (\ref{eq:4.13}), (\ref{eq:4.20i}-
\ref{eq:4.20ii}) and (\ref{eq:4.21}) we derive that:
\begin{equation}
\frac{\partial T^{\alpha\beta}(x)}{\partial x^{\beta}} = \rho
u^{\beta} \frac{\partial u^{\alpha}}{\partial x^{\beta}} +
u^{\alpha} \frac{\partial (\rho u^{\beta})}{\partial x^{\beta}} +
\frac{1}{c} F^{\lambda \alpha} J_{\lambda}. \label{eq:4.22}
\end{equation}
We postulate that a physical conservation law to hold:
\begin{equation}
\frac{\partial T^{\alpha\beta}}{\partial x^{\beta}}=0.
\label{eq:4.23}
\end{equation}
Moreover, we consider the case of a charged dust, i.e.
\begin{equation}
J^{\lambda}(x)=\sigma_{(0)}(x) U^{\lambda}(x). \label{eq:4.24}
\end{equation}
Here, $\sigma_{0}(x)$ is the proper charge density (the charge
density measured in the co-moving frame of the charge).  The
equation (\ref{eq:4.22}), with (\ref{eq:4.20ii}), yields
\begin{subequations}
\begin{align}
\frac{\partial}{\partial x^{\beta}}
\left[\rho(x)U^{\beta}(x)\right]=&0, \label{eq:4.25i} \\
\rho U^{\beta}(x) \frac{\partial U^{\alpha}(x)}{\partial x^{\beta}} =&
\frac{\sigma_{0}(x)}{c} U^{\lambda}(x) F^{\alpha}_{\;\; \lambda} .
\label{eq:4.25ii}
\end{align}
\end{subequations}
The equations (\ref{eq:4.25i}) stands for the continuity of the
material flow, whereas the equation (\ref{eq:4.25ii}) represents
the Lorentz equation (\ref{eq:4.14i}) for the charged dust.

\qquad Following the discussion on particle mechanics, we define
the three-dimensional velocity field components by:
\begin{align}
v^{i}=V^{i}(\mathbf{x},t):=&c\frac{U^{i}(x)}{U^{4}(x)}, \;\; U^{4}(x)=\frac{c}{\genrelgammax}, \nonumber \\
v^{i}=V^{i}(\mathbf{x},t)= &\genrelgammaxt \, U^{i}(x), \;\; \mbox{and
}\;\; J^{4}(x)=\sigma_{0}(x)U^{4}(x)=c\sigma(x). \label{eq:4.27}
\end{align}
The spatial components (\ref{eq:4.25ii}) provide:
\begin{equation}
\rho \left\{ \frac{\partial}{\partial t}
\left[\frac{v^{i}}{\genrelgammax}\right] +v^{k}
\frac{\partial}{\partial x^{k}} \left[\frac{v^{i}}{\genrelgammax}
\right] \right\} =\sigma_{0} \left[ E^{i} +
\epsilon^{ijk}\frac{v_{j}}{c} B_{k} \right]. \label{eq:4.28}
\end{equation}
Using notations of three-dimensional vector calculus, the
equation (\ref{eq:4.28}) can be cast into the more familiar form:
\begin{equation}
\rho \left\{ \frac{\partial}{\partial t}
\left[\frac{\mathbf{V}(\mathbf{x},t)}{\genrelgammaxt} \right]
+\mathbf{v} \cdot \mathbf{\nabla}
\left[\frac{\mathbf{V}(\mathbf{x},t)}{\genrelgammaxt} \right]
\right\} = \sigma_{0} \left[\mathbf{E} + \frac{\mathbf{v}}{c}
\times \mathbf{B} \right]. \label{eq:4.29}
\end{equation}

\section{Special relativistic gravitational fields}
\subsection{Introduction}
The static Newtonian gravitational potential, $W(x)$, satisfies:
\begin{subequations}
\begin{align}
\nabla^{2}W(\bfx):=& \delta^{ij} \frac{\partial^{2}
W(\bfx)}{\partial x^{i} \partial x^{j}}= \left\{
\begin{array}{ll}
4\pi G \rho(\bfx) & \mbox{inside a material body,}
 \\
0 & \mbox{in vacuum;}
\end{array}
\right. \label{eq:5.1i} \\
W(\bfx)=& -G \int_{body} \frac{\rho(\bfx^{\prime})}{|\bfx -
\bfx^{\prime}|}\,d^{3}x^{\prime}, \label{eq:5.1ii} \\
\rho(\mathbf{x}^{\prime}) \geq & 0, \;\;W(\mathbf{x}) \leq 0. \label{eq:5.1iii}
\end{align}
\end{subequations}
Here, $G$ is
the Newtonian constant of gravitation.

\qquad The well known equations of motion of a particle of $m >
0$ in the external gravitational field $W(\mathbf{x})$ are
furnished by:
\begin{subequations}
\begin{align}
m \frac{d\mathbf{V}(t)}{dt}=& -m
\mathbf{\nabla}W(\bfx)_{|\overline{\Chi}(t)}, \label{eq:5.2i} \\
\frac{dV^{i}(t)}{dt}=& - \frac{\partial W(\bfx)}{\partial
x^{i}}_{|\overline{\Chi}(t)} . \label{eq:5.2ii}
\end{align}
\end{subequations}

\qquad The natural relativistic generalization of the Newtonian
potential is a four-dimensional scalar field. Such a field has
been considered but the scalar field theory yields incorrect
planetary orbits \cite{ref:11}. Next in order of complication, we
should consider a four-dimensional vector field as in equation
(\ref{eq:4.19i}). However, material sources of such a field
\emph{repel} each other (like similar electric charges). Thus,
vector fields are ruled out as viable candidates for a
relativistic gravitational theory. Next in order of complication
is the four-dimensional second order tensor field. This type of
field is a potential candidate as the relativistic generalization
of the mass density (within a factor of $c^{2}$) is the energy
density. As discussed in the previous section, the energy density
is a component of a symmetric rank two tensor, the
energy-momentum-stress tensor. Therefore, it is natural to
consider a potential which is a symmetric second rank tensor
field with the energy-momentum-stress tensor as its source. The
first-order approximation of Einstein's general relativity theory
of gravity yields a second rank theory similar to that presented
below. The field equations, along with supplementary conditions
are summarized as the following:
\begin{subequations}
\begin{align}
\square \phi_{\mu\nu}(x) =& -2\kappa T_{\mu\nu}, \;\;\;\;\kappa:=
\frac{8\pi G}{c^{4}}, \label{eq:5.3i}
\\
\frac{\partial T^{\mu\nu}(x)}{\partial x^{\nu}} =&0,
\label{eq:5.3ii} \\
\frac{\partial \phi^{\mu\nu}(x)}{\partial x^{\nu}} =&0,
\label{eq:5.3iii} \\
\phi_{\mu\nu}(\bfx, x^{4})=&\frac{2G}{c^{4}} \int_{body}
\frac{T_{\mu\nu}\left(\bfx^{\prime},\, x^{4}-|\bfx -
\bfx^{\prime}|\right)}{|\bfx - \bfx^{\prime}|} \, d^{3}x^{\prime}
, \label{eq:5.3v}
\end{align}
\end{subequations}
Physically, the components of the tensor, $T_{\mu\nu}$, represent the
\emph{total density of energy, ($c$-times) momentum and stress} of the source
material. (A special example was furnished in the equation
(\ref{eq:4.21}).) The component $T_{44}(x)$, $T_{i4}(x)$ and
$T_{ij}(x)$ represent energy density, momentum density and stress
(or rate of stress) density respectively. The equation
(\ref{eq:5.3i}-\ref{eq:5.3iii}) are analogous to the
electromagnetic equations (\ref{eq:4.19i} - \ref{eq:4.19ii}) and
(\ref{eq:4.18}) respectively.

\qquad Now we shall express the equation of motion of a test
particle of mass $m > 0$ in a gravitational field. The equations,
though complicated, are aesthetically pleasing. Firstly, it is
convenient to introduce a related second order symmetric tensor
field by the definition:
\begin{align}
g_{\mu\nu}(x):=&d_{\mu\nu} + \phi_{\mu\nu}(x) -\frac{1}{2}
d_{\mu\nu}d^{\alpha\beta} \phi_{\alpha\beta}(x) \equiv
g_{\nu\mu}(x), \label{eq:5.4}
\end{align}
For weak gravitational fields, $|\phi_{\mu\nu}(x)| << 1$ and the
determinant of $g_{\mu\nu}(x)$ is close to $-1$. Thus, there
exists a unique inverse matrix defined by
\begin{align}
\left[g^{\mu\nu}(x) \right]:=& \left[g_{\mu\nu}(x)\right]^{-1},
\label{eq:5.5} \\
g^{\mu\nu}(x)g_{\nu\lambda}(x) = & \delta^{\mu}_{\;\lambda} .
\nonumber
\end{align}
Treating formally $g_{\mu\nu}(x)$ as a ``metric tensor'', we
define the associated Christoffel symbols as
\begin{equation}
\left\{
\begin{array}{l}
\alpha \\
\beta \; \gamma
\end{array}
\right\}:=\frac{1}{2} g^{\alpha\lambda}(x) \left[ \frac{\partial
g_{\gamma\lambda}(x)}{\partial x^{\beta}} + \frac{\partial
g_{\lambda\beta}(x)}{\partial x^{\gamma}} - \frac{\partial
g_{\beta\gamma}(x)}{\partial x^{\lambda}} \right] \equiv \left\{
\begin{array}{l}
\alpha \\
\gamma \; \beta
\end{array}\right\}.
\label{eq:5.6}
\end{equation}
In our linear theory, the above are components of a third order
four-dimensional tensor with forty independent components!

\qquad We postulate that the special relativistic equations of
motion, subject
only to gravity, must follow geodesic paths of the ``metric''
$g_{\mu\nu}(x)$:
\begin{equation}
\frac{d^{2}\Chi^{\alpha}(s)}{ds^{2}}+\left\{
\begin{array}{l}
\alpha \\
\beta \; \gamma
\end{array}
\right\}_{|\Chi(s)}\frac{d\Chi^{\beta}(s)}{ds}\frac{d\Chi^{\gamma}(s)}{ds} =0. \label{eq:5.7}
\end{equation}
Note that these equations of motion are independent of mass, a
property in common with the Newtonian theory. These semi-linear
equations distantly resemble the Lorentz equations
(\ref{eq:4.14i}) and generalize Newton's equations
(\ref{eq:5.2ii}) considerably. Geodesic equations (\ref{eq:5.7})
admit the exact first integral \cite{ref:7}
\begin{equation}
g_{\mu\nu}(x)_{|\Chi(s)}\frac{d\Chi^{\mu}(s)}{ds}\frac{d
\Chi^{\nu}(s)}{ds} \equiv \mbox{ constant}. \label{eq:5.8i}
\end{equation}
Since, in the limit $\phi_{\mu\nu}(x) \rightarrow 0$, the
equations (\ref{eq:5.8i}) reduce to (\ref{eq:3.18iv}) and
(\ref{eq:4.20ii}), we must choose
\begin{equation}
g_{\mu\nu}(x)_{|\Chi(s)}\frac{d\Chi^{\mu}(s)}{ds}\frac{d\Chi^{\nu}(s)}{ds} = -c^{2}. \label{eq:5.8ii}
\end{equation}

\qquad The generalization of (\ref{eq:5.7}) in the presence of
gravitational as well as non-gravitational force components
$\mathcal{F}^{\alpha}(x,u)$ is given by
\begin{subequations}
\begin{align}
m\frac{d^{2}\Chi^{\alpha}(s)}{ds^{2}}= -m\left\{
\begin{array}{l}
\alpha \\
\beta \; \gamma
\end{array}
\right\}_{|\Chi(s)}\frac{d\Chi^{\beta}(s)}{ds}\frac{d^{2}\Chi^{\gamma}(s)}{ds} +& \mathcal{F}^{\alpha}(x,u)_{|\Chi(s), d\Chi(s)/ds}, \label{eq:5.9i} \\
g_{\alpha\beta}(x)_{|\Chi(s)} \frac{d\Chi^{\alpha}(s)}{ds}
\mathcal{F}^{\beta}(x,u)_{|..} = & 0. \label{eq:5.9ii}
\end{align}
\end{subequations}
(Compare the above equations with (\ref{eq:3.24}) and (\ref{eq:3.25i})).

\qquad With the help of the ``metric'' tensor $g_{\mu\nu}(x)$, we
can define the familiar covariant derivative \cite{ref:7} and the
following consequences:
\begin{subequations}
\begin{align}
\nabla_{\alpha} A^{\beta}(x):=& \frac{\partial A^{\beta}(x)}{\partial x^{\alpha}} +\left\{
\begin{array}{l}
\beta \\
\alpha \; \gamma
\end{array}
\right\}A^{\gamma}(x) , \label{eq:5.10i} \\
\nabla_{\alpha} A_{\beta}(x):=& \frac{\partial A_{\beta}(x)}{\partial x^{\alpha}} -\left\{
\begin{array}{l}
\gamma \\
\alpha \; \beta
\end{array}
\right\}A_{\gamma}(x) , \label{eq:5.10ii} \\
\nabla_{\alpha} T_{\beta\gamma}(x):=& \frac{\partial T_{\beta\gamma}(x)}{\partial x^{\alpha}} -\left\{
\begin{array}{l}
\lambda \\
\alpha \; \beta
\end{array}
\right\}T_{\lambda\gamma}(x) -\left\{
\begin{array}{l}
\lambda \\
\alpha \; \gamma
\end{array}
\right\}T_{\alpha \lambda}(x), \label{eq:5.10iii} \\
\nabla_{\alpha}g_{\beta\gamma}(x) \equiv & 0, \;\; \nabla_{\alpha} g^{\beta\gamma}(x) \equiv 0, \;\; \nabla_{\alpha} \delta^{\beta}_{\;\gamma} \equiv 0, \label{eq:5.10iv} \\
\nabla_{\alpha}d_{\beta\gamma} \not\equiv& 0,\;\; \nabla_{\alpha}d^{\beta\gamma} \not\equiv 0. \label{eq:5.10v}
\end{align}
\end{subequations}

\subsection{Gravitational field of an incoherent dust}
\begin{figure}[ht]
\begin{center}
\includegraphics[bb=0 0 674 331, clip, scale=0.5, keepaspectratio=true]{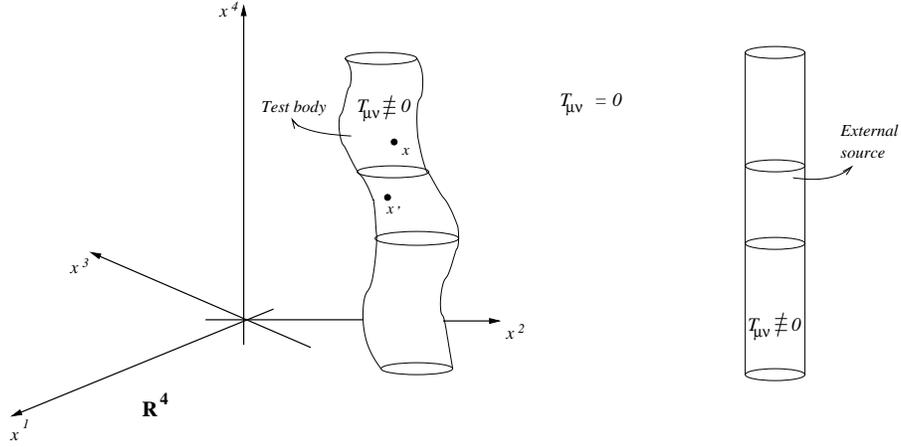}
\caption{{\small Gravitational forces on a test body due to an
external source in spacetime. The external source's $T_{\mu\nu}$
gives rise to $\phi_{(0)\mu\nu}$ while the test body's
$T_{\mu\nu}$ yields $\phi_{(I)\mu\nu}$. }} \label{fig:3}
\end{center}
\end{figure}

The gravitational equations (\ref{eq:5.3i} - \ref{eq:5.3iii}) are
all linear. Splitting the solution into its vacuum and
inhomogeneous parts we express:
\begin{subequations}
\begin{align}
\phi_{\mu\nu}(x)=& \phi_{(I)\mu\nu}(x) + \phi_{(0)\mu\nu}(x), \label{eq:5.11i} \\
\square \phi_{(I)\mu\nu}(x)=& -2\kappa T_{\mu\nu}(x), \;\; \frac{\partial T^{\mu\nu}(x)}{\partial x^{\nu}}=0, \;\;\frac{\partial \phi^{\mu\nu}_{(I)}(x)}{\partial x^{\nu}}=0, \label{eq:5.11ii} \\
\square \phi_{(0)\mu\nu}(x)=& 0, \;\; \frac{\partial \phi^{\mu\nu}_{(0)}(x)}{\partial x^{\nu}}=0, \label{eq:5.11iii} \\
\phi_{\mu\nu}(x)=& \phi_{(0)\mu\nu}(x) +\frac{2G}{c^{4}}
\int_{test\;\;body} \frac{T_{\mu\nu}
\left(\mathbf{x}^{\prime},\;x^{4}-|\mathbf{x}-\mathbf{x}^{\prime}|\right)}{|\mathbf{x}-\mathbf{x}^{\prime}|}\,
d^{3}x^{\prime}. \label{eq:5.11iv}
\end{align}
\end{subequations}
The above is the most general solution of the partial differential
equations (\ref{eq:5.3i}- \ref{eq:5.3iii}). Here
$\phi_{(I)\mu\nu}(x)$ represents the particular solution due to
the test body material with energy momentum stress tensor
$T_{\mu\nu}(x)$ and $\phi_{(0)\mu\nu}(x)$ represents the vacuum
solution created by any external sources (see figure
(\ref{fig:3})).

\qquad \emph{Ignoring} the internal field of the test body, we modify
(\ref{eq:5.4}) and (\ref{eq:5.8ii}) to
\begin{subequations}
\begin{align}
g_{(0)\mu\nu}:=d_{\mu\nu} + \phi_{(0)\mu\nu}(x) -&\frac{1}{2} d_{\mu\nu} d^{\alpha\beta} \phi_{(0)\alpha\beta}(x), \label{eq:5.12i} \\
g_{(0)\mu\nu}(x) U^{\mu}(x) U^{\nu}(x) \equiv & -c^{2}, \label{eq:5.12ii} \\
\overline{u}_{\mu}:=g_{(0)\mu\nu}u^{\nu}, \;\; \overline{u}_{\mu}u^{\mu} \equiv &-c^{2}. \label{eq:5.12iii}
\end{align}
\end{subequations}
For an incoherent dust coupled to gravity, the
energy-momentum-stress tensor is taken to be
\begin{equation}
T^{\mu\nu}(x):= \rho(x)U^{\mu}(x)U^{\nu}(x) + \Theta^{\mu\nu}(x,\rho, u^{\alpha}, \phi_{(0)\gamma\sigma}), \label{eq:5.13}
\end{equation}
Here $\rho(x)$ is the proper mass density, $U^{\mu}(x)$ is the
four velocity-field of the test particle and $\Theta^{\mu\nu}(x,
\rho, u^{\alpha}, \phi_{(0)\gamma\sigma})$ indicates the
\emph{interaction} of the test body with the external
gravitational field (from this point onward we will write this
term as $\Theta^{\mu\nu}(x)$).

\qquad The conservation equation (\ref{eq:5.3ii}) along with (\ref{eq:5.13}) yields:
\begin{align}
0=& \overline{u}_{\mu} \frac{\partial T^{\mu\nu}}{\partial x^{\nu}}= \overline{u}_{\mu} \left[u^{\mu} \frac{\partial (\rho u^{\nu})}{\partial x^{\nu}} + \rho u^{\nu} \frac{\partial u^{\mu}}{\partial x^{\nu}} + \frac{\partial \Theta^{\mu\nu}}{\partial x^{\nu}} \right] \nonumber \\
&= -c^{2} \frac{\partial (\rho u^{\nu})}{\partial x^{\nu}} + \rho u^{\nu} \overline{u}_{\mu} \left[\frac{\partial u^{\mu}}{\partial x^{\nu}} + \left\{
\begin{array}{l}
\mu \\
\nu \; \lambda
\end{array}
\right\}_{(0)} u^{\lambda} \right] + \overline{u}_{\mu} \left[
\frac{\partial \Theta^{\mu\nu}}{\partial x^{\nu}} -\rho \left\{
\begin{array}{l}
\lambda \\
\nu\; \lambda
\end{array}
\right\}_{(0)} u^{\nu}u^{\lambda} \right]. \label{eq:5.14}
\end{align}
Note that the middle terms in the above equation vanishes via
\begin{align}
&\overline{u}_{\mu} \left[ \frac{\partial u^{\mu}}{\partial
x^{\nu}} +
\left\{
\begin{array}{l}
\mu \\
\nu\; \lambda
\end{array}
\right\}_{(0)}
u^{\lambda} \right] \nonumber \\
&=\overline{u}_{\mu} \nabla_{(0)\nu} u^{\mu}= \frac{1}{2}
\nabla_{(0)\nu} \left[ g_{(0)\mu\nu}u^{\nu}u^{\mu}\right]
=-\frac{1}{2}\nabla_{(0)\nu}(c^{2})\equiv 0. \label{eq:90ii}
\end{align}
At this stage there are several possible choices to enforce the
conservation law (\ref{eq:5.14}). We stipulate the condition:
\begin{equation}
\frac{\partial \Theta^{\mu\nu}}{\partial x^{\nu}}= \rho \left\{
\begin{array}{l}
\lambda \\
\alpha \; \beta
\end{array}
\right\}_{(0)} u^{\nu}u^{\lambda}, \label{eq:5.15}
\end{equation}
as this choice yields the correct physical laws as dictated by the continuity equation \cite{ref:12}:
\begin{equation}
\frac{\partial}{\partial x^{\nu}} \left[\rho(x) U^{\nu}(x)\right] =0. \label{eq:5.16}
\end{equation}
The right hand side of (\ref{eq:5.15}) therefore represents the gravitational force components due to external sources.

\qquad With the help of equations (\ref{eq:4.27}), we can express (\ref{eq:5.16}) in a slightly more familiar form:
\begin{equation}
\frac{\partial}{\partial t} \left[\frac{\rho}{\flgenrelgamma} \right] + \mathbf{\nabla}\cdot \left[ \frac{\rho \mathbf{v}}{\flgenrelgamma} \right] =0. \label{eq:5.16ii}
\end{equation}

Substituting (\ref{eq:5.15}) and (\ref{eq:5.16}) into the
conservation equation (\ref{eq:5.3ii}) with (\ref{eq:5.10i}) and
(\ref{eq:5.13}) we obtain
\begin{equation}
u^{\nu} \left[ \frac{\partial u^{\mu}}{\partial x^{\nu}} + \left\{
\begin{array}{l}
\mu \\
\nu \; \lambda
\end{array}
\right\}_{(0)} u^{\lambda} \right] =u^{\nu} \nabla_{(0)\nu} u^{\mu} =0. \label{eq:5.17}
\end{equation}
For the streamlines, we need to solve the system of first-order ordinary differential equations:
\begin{equation}
\frac{d \Chi^{\mu}(s)}{ds} = U^{\alpha}(x)_{|\Chi(s)}.
\label{eq:5.18}
\end{equation}
(We assume the validity of the Lifshitz conditions \cite{ref:13}
on the right-hand-side of (\ref{eq:5.18}).) Using the above
equation in (\ref{eq:5.17}) we derive that
\begin{equation}
\frac{d^2 \Chi^{\mu}(s)}{ds^{2}} + \left\{
\begin{array}{l}
\mu \\
\nu \; \lambda
\end{array}
\right\}_{(0)|\Chi(s)} \frac{d\Chi^{\nu}(s)}{ds}\frac{d\Chi^{\lambda}(s)}{ds}=0 . \label{eq:5.19}
\end{equation}
Therefore, streamlines \emph{follow ``geodesics'' of the external
metric} $g_{(0)\mu\nu}(x)$. The Christoffel symbols mediate the
external gravitational forces on a test particle.

\subsection{Gravitational field of a charged dust}
A fluid with vanishing pressure is know as a ``dust''. In this case, the combined electromagnetic and gravitational field
equations are investigated \cite{ref:14}. Using (\ref{eq:4.10i} -
\ref{eq:4.10iii}), (\ref{eq:4.24}), \ref{eq:4.12i}),
(\ref{eq:4.20i}) and (\ref{eq:5.11i}- \ref{eq:5.11iii}), we
obtain the pertinent equations to be:
\begin{subequations}
\begin{align}
\frac{\partial F^{\mu\nu}}{\partial x^{\nu}} =&\frac{4\pi}{c} \sigma_{(0)} u^{\mu}, \label{eq:5.20i} \\
\frac{\partial F^{\mu\nu}}{\partial x^{\lambda}} + \frac{\partial F^{\nu\lambda}}{\partial x^{\mu}} + \frac{\partial F^{\lambda\mu}}{\partial x^{\nu}}=& 0, \label{eq:5.20ii} \\
\frac{\partial (\sigma_{(0)} u^{\nu})}{\partial x^{\nu}} =&0, \label{eq:5.20iii} \\
\phi_{\alpha\beta}(x) =& \phi_{(I) \alpha\beta}(x) +\phi_{(0)\alpha\beta}(x), \label{eq:5.20iv} \\
\square \phi_{(0) \alpha\beta}=\frac{\partial \phi_{(0)}^{\alpha\beta}}{\partial x^{\beta}} =&0, \label{eq:5.20v} \\
\square \phi_{(I)\alpha\beta}=-2 \kappa T_{\alpha\beta}:=& -2\kappa\left[\rho u_{\alpha}u_{\beta} + \Theta_{\alpha\beta} + \mathcal{M}_{\alpha\beta} \right], \label{eq:5.20vi} \\
\frac{\partial T^{\alpha\beta}}{\partial x^{\beta}}=&0, \label{eq:5.20vii} \\
\frac{\partial \phi^{\alpha\beta}_{(I)}}{\partial x^{\beta}}=&0. \label{eq:5.20viii}
\end{align}
\end{subequations}
\qquad Using (\ref{eq:5.20vi}), (\ref{eq:4.24}) and
(\ref{eq:4.22}), the conservation equation (\ref{eq:5.20vii})
yields:
\begin{equation}
0=\frac{\partial T^{\alpha\beta}}{\partial x^{\beta}}=u^{\alpha} \frac{\partial(\rho u^{\beta})}{\partial x^{\beta}} + \rho u^{\beta} \frac{\partial u^{\alpha}}{\partial x^{\beta}} +\frac{\partial \Theta^{\alpha\beta}}{\partial x^{\beta}} + \frac{\sigma_{(0)}}{c} u^{\beta} F^{\;\;\alpha}_{\beta}. \label{eq:5.21i}
\end{equation}

\qquad Imposing the condition (\ref{eq:5.15}) on
$\Theta^{\alpha\beta}(x)$, the above equation (\ref{eq:5.21i})
leads to
\begin{equation}
u^{\alpha} \frac{\partial(\rho u^{\beta})}{\partial x^{\beta}} +\rho u^{\beta} \nabla^{(0)}_{\beta} u^{\alpha} +\frac{\sigma_{(0)}}{c} u^{\beta} F^{\;\;\alpha}_{\beta} =0. \label{eq:5.21ii}
\end{equation}
Now, the equations (\ref{eq:5.10iv}) and (\ref{eq:5.12ii})
provide:
\begin{align}
2\overline{u}_{\alpha} \nabla^{(0)}_{\beta} u^{\alpha}=&
g_{(0)\alpha\gamma} \left[u^{\gamma} \nabla^{(0)}_{\beta}
u^{\alpha} +u^{\alpha} \nabla^{(0)}_{\beta} u^{\gamma} \right]
\nonumber \\ =&\nabla^{(0)}_{\beta} \left[g_{(0)\alpha\gamma}
u^{\alpha}u^{\gamma} \right] = \nabla^{(0)}_{\beta}
\left(-c^{2}\right) \equiv 0. \label{eq:5.22}
\end{align}
($\nabla^{(0)}_{\beta}$ is the covariant derivative defined with
respect to the metric $g_{(0)\mu\nu}$, defined in
(\ref{eq:5.12i}).) Therefore, (\ref{eq:5.22}), (\ref{eq:5.14})
and (\ref{eq:5.21ii}) yield
\begin{equation}
\frac{\partial (\rho u^{\beta})}{\partial x^{\beta}} = \frac{\sigma_{(0)}}{c^{3}} \overline{u}_{\alpha} u^{\beta} F_{\beta}^{\;\;\alpha} . \label{eq:5.23}
\end{equation}
Substituting (\ref{eq:5.23}) into (\ref{eq:5.21ii}), we finally obtain
\begin{equation}
\rho u^{\beta} \nabla^{(0)}_{\beta} u^{\alpha}
+\frac{\sigma_{(0)}}{c}u^{\beta} F^{\;\;\gamma}_{\beta}
\left[\delta^{\alpha}_{\;\gamma} +\frac{1}{c^{2}}
\overline{u}_{\gamma}u^{\alpha}  \right]=0. \label{eq:5.24}
\end{equation}
Hence, the stream lines of a charged dust satisfying
(\ref{eq:5.18}) must pursue trajectories governed by the
equations of motion:
\begin{align}
&\left\{ \rho(x) \left[ \frac{d^{2} \Chi^{\alpha}(s)}{ds^{2}} +
\left\{
\begin{array}{l}
\alpha \\
\beta \; \gamma
\end{array}
\right\}_{(0)}
\frac{d\Chi^{\beta}(s)}{ds}\frac{d \Chi^{\gamma}(s)}{ds}\right] \right. \nonumber \\
& \left. +\frac{\sigma_{(0)}(x)}{c} F^{\;\;\gamma}_{\beta}(x)
\left[\delta^{\alpha}_{\;\gamma}+\frac{1}{c^{2}} g_{(0)\gamma\mu}
\frac{d \Chi^{\alpha}(s)}{ds}\frac{d
\Chi^{\mu}(s)}{ds}\right]\frac{d\Chi^{\beta}(s)}{ds}
\right\}_{|x^{\alpha}=\Chi^{\alpha}(s)} =0. \label{eq:5.25}
\end{align}
These are the modified relativistic Lorentz equations of motion
in the presence of external gravitational fields. (Compare with
the equations (\ref{eq:4.25ii}.) and (\ref{eq:4.28}).)

\subsection{Gravitational field of a perfect fluid}
Following the prescription in the previous sections, a special
relativistic perfect fluid in a gravitational field is governed by
\begin{subequations}
\begin{align}
\phi_{\alpha\beta}(x) =&\phi_{(I)\alpha\beta}(x) +
\phi_{(0)\alpha\beta}(x), \label{eq:5.26i} \\
\square \phi_{(0)\alpha\beta}(x)=& 0= \frac{\partial
\phi_{(0)}^{\alpha\beta}}{\partial x^{\beta}},
\label{eq:5.26ii} \\
\square \phi_{(I)}^{\alpha\beta}(x)=& -2\kappa T^{\alpha\beta}(x)
\nonumber \\
 :=& -2\kappa \left[\left(\rho(x) +\frac{p(x)}{c^{2}}
\right) u^{\alpha}u^{\beta} +p(x) g_{(0)}^{\alpha\beta}(x) +
\Theta^{\alpha\beta}(x)\right], \label{eq:5.26iii} \\
\frac{\partial T^{\alpha\beta}}{\partial x^{\beta}}=&0,
\label{eq:5.26iv} \\
\frac{\partial \phi^{\alpha\beta}_{(I)}}{\partial x^{\beta}}=&0.
\label{eq:5.26v}
\end{align}
\end{subequations}
Here, $p(x)$ is the pressure.

\qquad The conservation equation (\ref{eq:5.26iv}) implies from
(\ref{eq:5.26iii}) that
\begin{align}
0=& \frac{\partial T^{\alpha\beta}}{\partial x^{\beta}}=
u^{\alpha} \frac{\partial}{\partial x^{\beta}} \left[\left(\rho
+\frac{p}{c^{2}}\right)u^{\beta}\right] +\left(\rho
+\frac{p}{c^{2}}\right) u^{\beta} \nabla^{(0)}_{\beta} u^{\alpha}
+\frac{\partial}{\partial x^{\beta}} \left[pg_{(0)}^{\alpha\beta}
\right] \nonumber \\
&+ \left[ \frac{\partial \Theta^{\alpha\beta}}{\partial
x^{\beta}} - \left(\rho +\frac{p}{c^{2}} \right)\left\{
\begin{array}{l}
\alpha \\
\beta \; \gamma
\end{array}
\right\}_{(0)} u^{\beta} u^{\gamma} \right]. \label{eq:5.27}
\end{align}
Following the previous section, reasonable physics demands that we stipulate
\begin{equation}
\frac{\partial \Theta^{\alpha\beta}}{\partial x^{\beta}}
=\left(\rho +\frac{p}{c^{2}}\right)
\left\{
\begin{array}{l}
\alpha \\
\beta \; \gamma
\end{array}
\right\}_{(0)}
u^{\beta}u^{\gamma}.
\label{eq:5.28}
\end{equation}
Note that the right-hand-side denotes the external gravitational
forces on the \emph{effective mass density} $\left(\rho
+\frac{p}{c^{2}}\right)$. Now, the equation (\ref{eq:5.27}) yields
\begin{equation}
u^{\alpha} \frac{\partial}{\partial x^{\beta}} \left[\left(\rho
+\frac{p}{c^{2}}\right)u^{\beta}\right] +\left(\rho
+\frac{p}{c^{2}}\right) u^{\beta} \nabla^{(0)}_{\beta} u^{\alpha}
= -\frac{\partial}{\partial x^{\beta}}
\left[pg_{(0)}^{\alpha\beta} \right]. \label{eq:5.29}
\end{equation}
Contracting this equation with $\overline{u}_{\alpha}$ and using
(\ref{eq:5.12ii}) and (\ref{eq:5.22}), we deduce the continuity
equation,
\begin{equation}
\frac{\partial}{\partial x^{\beta}} \left[\left(\rho
+\frac{p}{c^{2}} \right) u^{\beta} \right] =\frac{1}{c^{2}}
\overline{u}_{\alpha} \frac{\partial}{\partial x^{\beta}} \left(p
g^{\alpha\beta}_{(0)} \right). \label{eq:5.30}
\end{equation}
Using this equation in (\ref{eq:5.29}), we derive the
relativistic Euler equation
\begin{equation}
\left(\rho +\frac{p}{c^{2}} \right) u^{\beta}
\nabla^{(0)}_{\beta} u^{\alpha} =-\left[\delta^{\alpha}_{\;\gamma}
+\frac{\overline{u}_{\gamma}u^{\alpha}}{c^{2}} \right]
\frac{\partial}{\partial x^{\beta}} \left(p
g_{(0)}^{\gamma\beta}\right). \label{eq:5.31}
\end{equation}

\qquad On a typical stream line (given by (\ref{eq:5.18})) of the
perfect fluid, the following equations of motion hold:
\begin{align}
&\left[\rho +\frac{p}{c^{2}} \right]
\left[\frac{d^{2}\Chi^{\alpha}(s)}{ds^{2}} +\left\{
\begin{array}{l}
\alpha \\
\beta \; \gamma
\end{array}
\right\}_{(0)} \frac{d \Chi^{\beta}(s)}{ds} \frac{d
\Chi^{\gamma}(s)}{ds} \right]_{|x^{\alpha}=\Chi^{\alpha}(s)}
\nonumber \\
&= - \left[\delta^{\alpha}_{\;\gamma} +\frac{1}{c^{2}}
g_{(0)\gamma\mu} \frac{d \Chi^{\alpha}(s)}{ds} \frac{d
\Chi^{\mu}(s)}{ds} \right]\left\{\frac{\partial}{\partial
x^{\beta}} \left[p g^{\gamma\beta}_{(0)} \right]
\right\}_{|x^{\alpha}=\Chi^{\alpha}(s)}. \label{eq:5.32}
\end{align}

\section{Static external gravitational fields}
The special relativistic gravitaitonal fields may easily be
compared with Newtonian theory in this case. The relevant
relativistic equations read:
\begin{subequations}
\begin{align}
\phi_{(0)ij}(x) \equiv & 0,\;\; \phi_{(0) i4}(x) \equiv 0, \label{eq:6.1i} \\
\phi_{(0)44}(x)\neq &0, \;\; \frac{\partial \phi_{(0)44}(x)}{\partial x^{4}} \equiv 0, \label{eq:6.1ii} \\
d^{\alpha\beta} \phi_{(0)\alpha\beta}(x)=&
-\phi_{(0)44}(\mathbf{x}),
\label{eq:6.1iii} \\
g_{(0)ij}(x)=& \left[1+\frac{1}{2} \phi_{(0)44}(\mathbf{x}) \right] \delta_{ij}, \label{eq:6.1iv} \\
g_{(0)i4}(x) \equiv & 0, \;\; g_{(0)44}(x)=-1+ \frac{1}{2}
\phi_{(0)44}(\mathbf{x}) .\label{eq:6.1v}
\end{align}
\end{subequations}
(We have made use of equations (\ref{eq:5.12i}) to derive the above.)

\qquad The equations (\ref{eq:5.11iii}) reduce Laplace's equation:
\begin{equation}
\nabla^{2} \phi_{(0)44}(\mathbf{x}) =0. \label{eq:6.2}
\end{equation}
Comparing (\ref{eq:6.2}) with (\ref{eq:5.1i}) ,
(\ref{eq:5.1ii}) and (\ref{eq:5.11iv}) we identify $\phi_{(0)44}(\mathbf{x})$ with the
Newtonian gravitational potential via
\begin{equation}
\phi_{(0)44}(\mathbf{x}) =- \frac{4}{c^{2}} W(\mathbf{x}) = \frac{4}{c^{2}} |W(\mathbf{x})| \geq 0. \label{eq:6.3}
\end{equation}
We also note that for a four-dimensional vector field $T^{\alpha}(x)$, the equations (\ref{eq:6.1iv}), (\ref{eq:6.1v}) and (\ref{eq:6.3}) yield
\begin{align}
-g_{(0)\alpha\beta}(x) T^{\alpha}(x) T^{\beta}(x)&= \left[1-\frac{1}{2} \phi_{(0)44}(\mathbf{x})\right]\left[T^{4}(x)\right]^{2} -\left[1+ \frac{1}{2} \phi_{(0)44}(\mathbf{x}) \right] \delta_{ij} T^{i}(x) T^{j}(x) \nonumber \\
=&\left[1-\frac{2|W(\mathbf{x})|}{c^{2}}\right]\left[T^{4}(x)\right]^{2} -\left[1+ \frac{2 |W(\mathbf{x})|}{c^{2}} \right] \delta_{ij} T^{i}(x) T^{j}(x). \label{eq:6.4}
\end{align}
We have implicitly assumed that $2\frac{|W(\mathbf{x})|}{c^{2}} < 1$. The \emph{physical} components of the vector field $T^{\alpha}(x)$ are
\begin{align}
\overline{T}^{i}(x):=&\sqrt{\plusfact}T^{i}(x), \nonumber \\
\overline{T}^{4}(x):=&\sqrt{\minusfact}T^{4}(x). \label{eq:113ii}
\end{align}
Notice that these components are just the components of $T^{\alpha}(x)$ projected into the corresponding \emph{orthonormal} coordinates (or frame) of the metric $g_{(0)\mu\nu}$.

\subsection{Test particle motions in external static gravitational fields}
In a previous section it was noted that any stream line of incoherent dust follows the ``geodesic'' equation (\ref{eq:5.19}). Along any of these time-like geodesics, the equation (\ref{eq:5.8ii}) implies that
\begin{equation}
g_{(0)\mu\nu|\Chi(s)} \frac{d\Chi^{\mu}(s)}{ds}\frac{d\Chi^{\nu}(s)}{ds} \equiv -c^{2}. \label{eq:6.5}
\end{equation}
The above condition modifies  the equation (\ref{eq:3.18iv}). For the sake of consistency, we must alter the definition (\ref{eq:3.15}) for the proper time along a time-like curve by:
\begin{subequations}
\begin{align}
s=& S^{\#}(u):=\frac{1}{c} \int_{u_{1}}^{u} \sqrt{ -g_{(0)\mu\nu}(x)_{|..} \frac{d\Chi^{\#\,\mu}(w)}{dw} \frac{d\Chi^{\#\,\nu}(w)}{dw}}\,dw, \label{eq:6.6i} \\
s=& \overline{S}(t):=\frac{1}{c} \int_{t_{1}}^{t} \sqrt{ -g_{(0)\mu\nu}(x)_{|..} \frac{d\overline{\Chi}^{\mu}(w)}{dw} \frac{d\overline{\Chi}^{\nu}(w)}{dw}}\,dw, \label{eq:6.6ii} \\
s=& S(s):=\frac{1}{c} \int_{0}^{s} \sqrt{ -g_{(0)\mu\nu}(x)_{|..}
\frac{d\Chi^{\mu}(w)}{dw} \frac{d\Chi^{\nu}(w)}{dw}}\,dw.
\label{eq:6.6iii}
\end{align}
\end{subequations}

\qquad The equation (\ref{eq:3.18iv}) is modified by (\ref{eq:6.4}) and (\ref{eq:6.5}) into
\begin{equation}
\left[1+\frac{2|W(\mathbf{x})|}{c^{2}}\right]_{|..}\delta_{ij} \frac{d\Chi^{i}(s)}{ds} \frac{d\Chi^{j}(s)}{ds} -\left[1-\frac{2|W(\mathbf{x})|}{c^{2}}\right]_{|..} \left[\frac{d\Chi^{4}(s)}{ds}\right]^{2} \equiv -c^{2} < 0, \label{eq:6.7}
\end{equation}
while the equation (\ref{eq:3.22}) is changed by (\ref{eq:6.7}) into
\begin{equation}
\frac{ds}{dt}= \frac{d\overline{S}(t)}{dt} =\sqrt{1- \frac{2|W(\mathbf{x})|}{c^{2}} - \left[ 1+ \frac{2|W(\mathbf{x})|}{c^{2}} \right] \frac{|\mathbf{V}(t)|^{2}}{c^{2}}}_{\;\;|..} \leq 1. \label{eq:6.8}
\end{equation}
The above equation reveals the \emph{time dilation} along a moving dust particle in an external gravitational field.

\qquad Similarly, the equations (\ref{eq:3.23ii}) and (\ref{eq:3.23iii}) change over into:
\begin{subequations}
\begin{align}
U^{i}(s)= \frac{d\Chi^{i}(s)}{ds} =&\frac{V^{i}(t)}{\sqrt{\minusfact
-\left(\plusfact\right)\frac{|\mathbf{V}|^{2}}{c^{2}}}}_{\;|..}, \label{eq:6.9i} \\
U^{4}(s)= \frac{d\Chi^{4}(s)}{ds} =& \frac{c}{\sqrt{\minusfact
-\left(\plusfact\right)\frac{|\mathbf{V}|^{2}}{c^{2}}}}_{\;|..}.
\label{eq:6.9ii}
\end{align}
\end{subequations}
Here, $V^{i}(t) = \frac{d\overline{\Chi}^{i}(t)}{dt}$ are the
components of the ``coordinate'' velocity which in general differ
from the ``measurable'' velocity components,
$\overline{V}^{i}(t):= \sqrt{\frac{\plusfact}{\minusfact}}
\frac{d\overline{\Chi}^{i}(t)}{dt}_{\;|..}$.

\qquad Now, the geodesic equations (\ref{eq:5.19}) are derivable
from a variational principle \cite{ref:7}. A free particle's
motion is governed by a purely kinetic Lagrangian, $\frac{1}{2}m\,
\delta_{ij}V^{i}V^{j}$ in non-relativistic mechanics. The
relativistic analogue is $\frac{1}{2}m\,
g_{(0)\alpha\beta}u^{\alpha}\overline{u}^{\beta}$. In the presence
of gravity, the gravitational coupling arises from lowering the
index with the metric $g_{(0)\alpha\beta}$ and therefore, using
(\ref{eq:6.1iv}), (\ref{eq:6.1v}) and (\ref{eq:6.3}) the
Lagrangian (per unit mass) becomes
\begin{align}
L(x,u):=& \frac{1}{2} \left[\plusfact\right] \delta_{ij} u^{i}u^{j} -\frac{1}{2} \left[\minusfact\right] (u^{4})^{2} \nonumber \\
=& \frac{1}{2} \delta_{ij} u^{i}u^{j} -W(\mathbf{x}) \left(\frac{u^{4}}{c}\right)^{2} - \frac{(u^{4})^{2}}{2} -\frac{W(\mathbf{x})}{c^{2}} \delta_{ij}u^{i}u^{j}. \label{eq:6.10i}
\end{align}
The relativistic Euler-Lagrange equations are given by
\begin{equation}
\frac{\partial L(x,u)}{\partial x^{\alpha}}_{|x^{\alpha}=\Chi^{\alpha}(s), u^{\alpha}= \frac{d\Chi^{\alpha}(s)}{ds}} - \frac{d}{ds} \left[\frac{\partial L(x,u)}{\partial u^{\alpha}}_{|..}\right] =0. \label{eq:6.10ii}
\end{equation}

\qquad The Newtonian Lagrangian for the corresponding Newtonian theory is
\begin{equation}
L_{N}(\mathbf{x}, \mathbf{v}):= \frac{1}{2} \delta_{ij} v^{i}v^{j} -W(\mathbf{x}), \label{eq:6.11}
\end{equation}
giving rise to the equations of motion (\ref{eq:5.2ii}).

\qquad For small velocities, by the equations (\ref{eq:6.9i}), (\ref{eq:6.10i}) matches (\ref{eq:6.11}) except for the term $-(u^{4})^{2}/2$. This term represents the large rest energy contribution which is manifest in the relativistic physics.

\qquad It will be instructive to investigate the fourth equation
of (\ref{eq:6.10ii}). Since $x^{4}$ is a cyclic variable in the
Lagrangian (\ref{eq:6.10i}), the corresponding equation of motion
admits the first integral:
\begin{subequations}
\begin{align}
&\frac{\partial L(x,u)}{\partial u^{4}}_{|..} =-\left[\plusfactnabs\right]U^{4}(s)= const. , \label{eq:6.12i} \\
\mbox{or,}\;\;\; &\frac{c \left[\plusfactnabs\right]}{\sqrt{\plusfactnabs -\frac{1}{c^{2}} \left[\minusfactnabs \right] \delta_{ij}V^{i}V^{j}}}_{|..} =\frac{\mathcal{E}}{c} =const. \label{eq:6.12ii}
\end{align}
\end{subequations}
For \emph{small} velocities and \emph{weak} gravitational fields,
equation (\ref{eq:6.12ii}) yields:
\begin{equation}
\mathcal{E}=c^{2} + \left[ \frac{1}{2} |\mathbf{V}|^{2}
+W(\mathbf{x})\right]_{|..}
+\mathcal{O}\left(\frac{1}{c^{2}}\right). \label{eq:6.13}
\end{equation}

\qquad It is not difficult to interpret the above equation. The
constant $\mathcal{E}$ represents the conserved total energy per
unit mass. The first term on the RHS is the large rest energy of
the unit mass. The second and third terms are the usual kinetic
and potential energies respectively.

\subsubsection{An example from planetary motion}
Here we investigate planetary motions due to the spherically
symmetric gravitational field of the sun. The usual potential
function is furnished by:
\begin{equation}
W(\mathbf{x})= \frac{-GM}{\sqrt{(x^{1})^{2} +(x^{2})^{2} +
(x^{3})^{2}}}, \label{eq:6.14}
\end{equation}
with $M > 0$ being the solar mass. Employing spherical polar
coordinates in space, the equations (\ref{eq:6.10i}) and
(\ref{eq:6.14}) yield the Lagrangian
\begin{equation}
L(x,u)= \frac{1}{2} \left(1+\frac{2GM}{c^{2}r}\right) \left[ (u^{r})^{2} + r^{2} (u^{\theta})^{2} + r^{2}\sin^{2}\theta (u^{\varphi})^{2}\right] - \frac{1}{2} \left( 1-\frac{2GM}{c^{2}r}\right) (u^{4})^{2} . \label{eq:6.15}
\end{equation}

\qquad The Euler-Lagrange equations (\ref{eq:6.10i}), from (\ref{eq:6.15}) admit uniplanar motions characterized by:
\begin{equation}
\theta=\Theta(s) \equiv \frac{\pi}{2}. \label{eq:6.16}
\end{equation}
The reduced Lagrangian is then
\begin{equation}
L_{(0)}(..)=\frac{1}{2} \left(1+\frac{2GM}{c^{2}r}\right) \left[(u^{r})^{2} +r^{2}(u^{\varphi})^{2} \right] -\frac{1}{2} \left(1-\frac{2GM}{c^{2}r}\right)(u^{4})^{2} . \label{eq:6.17i}
\end{equation}
In this case, the conservation of energy equation (\ref{eq:6.12i}) reads:
\begin{equation}
\left[1 -\frac{2GM}{c^{2}r} \right] U^{4}(s)=\frac{\mathcal{E}}{c} . \label{eq:6.17}
\end{equation}

\begin{figure}[ht]
\begin{center}
\includegraphics[bb=0 0 185 211, clip, scale=0.5, keepaspectratio=true]{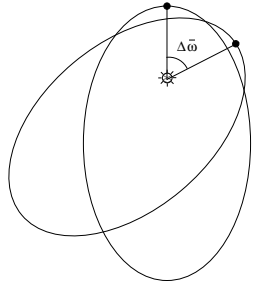}
\caption{{\small The perihelion precession of the planets in a gravitational field. The diagram displays two orbits, perhaps millenia apart, of a planet around a star. The perihelion point of the planet has shifted by an amount $\Delta\overline{\omega}$, an effect not predicted by Newtonian gravity theory.}} \label{fig:perihelion}
\end{center}
\end{figure}

\qquad The presence of another cyclic coordinate, $\varphi$, leads to the conservation of angular momentum:
\begin{equation}
\frac{\partial L_{(0)}(..)}{\partial u_{\varphi}}_{|..} =r^{2}
\left(1+\frac{2GM}{c^{2}r} \right) u^{\varphi}_{\;\; |..} = h =
const. \label{eq:6.18}
\end{equation}
Substituting (\ref{eq:6.16}), (\ref{eq:6.17}) and (\ref{eq:6.18})
into (\ref{eq:6.7}), we obtain
\begin{equation}
\left[ \left(1+\frac{2GM}{c^{2}r}\right)(u^{r})^{2}
+\frac{h^{2}}{r^{2}\left(1+\frac{2GM}{c^{2}r} \right)
}-\frac{\mathcal{E}^{2}}{c^{2} \left(1-\frac{2GM}{c^{2}r}\right)}
\right]_{|..} \equiv -c^{2}. \label{eq:6.19}
\end{equation}
We reparameterize the curve by the following:
\begin{align}
r=&\mathcal{R}(s) =\hat{\mathcal{R}}(\varphi), \label{eq:6.20} \\
\frac{d\mathcal{R}(s)}{ds}=& u^{\varphi}_{\;\; |..}\frac{d \hat{\mathcal{R}}(\varphi)}{d\varphi} =\frac{h}{r^{2} \left(1+\frac{2GM}{c^{2}r}\right)} \frac{d \hat{\mathcal{R}}(\varphi)}{d\varphi}.  \nonumber
\end{align}
It is useful at this point to make the following coordinate transformation:
\begin{align}
y=& \frac{1}{r},\;\; y=Y(\phi), \label{eq:6.21} \\
y^{\prime}:=& \frac{dY(\varphi)}{d\varphi} . \nonumber
\end{align}
With (\ref{eq:6.20}) and (\ref{eq:6.21}), the equation (\ref{eq:6.19}) reduces to
\begin{equation}
\left[ \frac{1-\frac{2GMy}{c^{2}}}{1+\frac{2GMy}{c^{2}}} \right]
\left[ (y^{\prime})^{2} +y^{2} \right] +\frac{c^{2}}{h^{2}}
\left( 1-\frac{2GMy}{c^{2}}\right)
=\frac{\mathcal{E}^{2}}{h^{2}c^{2}}. \label{eq:6.22}
\end{equation}
The above first-order equation can be solved by quadrature. However, to extract physically important effects, we differentiate the equation (\ref{eq:6.22}) to get
\begin{align}
y^{\prime\prime}+y=& \left(1+\frac{4GMy}{c^{2}}\right) \left[\frac{GM}{h^{2}} +\frac{2GM}{c^{2}} \left((y^{\prime})^{2} +y^{2} \right)\right] +\mathcal{O} \left(\frac{1}{c^{4}}\right) \nonumber \\
=& \frac{GM}{h^{2}} + 4 \left(\frac{GM}{hc}\right)^{2}y +\frac{2GM}{c^{2}} \left((y^{\prime})^{2} +y^{2} \right) +\mathcal{O} \left(\frac{1}{c^{4}}\right). \label{eq:6.23}
\end{align}
The equation (\ref{eq:6.23}) may be solved by the perturbative expansion:
\begin{equation}
y=y_{0} +\frac{y_{1}}{c^{2}} + \frac{y_{2}}{c^{4}} +... \label{eq:6.24}
\end{equation}
Using this expansion in (\ref{eq:6.23}), we obtain
\begin{equation}
y=y_{0} + \frac{y_{1}}{c^{2}}+...= \frac{GM}{h^{2}} \left[1+e\cos(\varphi -\overline{\omega})\right] +\frac{4(GM)^{3}}{c^{2}h^{4}} e\varphi \sin(\varphi-\overline{\omega}) +... \;\;. \label{eq:6.25}
\end{equation}
Here, the constants of integration $e$ and $\overline{\omega}$
represent the eccentricity and perihelion angle of the orbit
respectively. Combining the first two terms in (\ref{eq:6.25}),
we conclude that
\begin{equation}
y_{0} +\frac{y_{1}}{c^{2}} =\frac{GM}{h^{2}} \left[1+e\cos (\varphi -\overline{\omega} -\delta \overline{\omega}) \right] +\left(\mbox{higher order}\right) , \label{eq:6.26}
\end{equation}
where
\begin{equation}
\delta\overline{\omega}:=\arctan\left[4\varphi \left(\frac{GM}{h}\right)^{2} \right] =4 \left(\frac{GM}{h}\right)^{2}\varphi + \left(\mbox{higher order}\right). \nonumber
\end{equation}
Thus, the elliptic orbit precesses and the perihelion angle changes (see figure \ref{fig:perihelion}). This gravitational perihelion shift per revolution is given by
\begin{equation}
\Delta \overline{\omega}=8\pi \left(\frac{GM}{h}\right)^{2}. \label{eq:6.27}
\end{equation}
For the planet Mercury, the above amount yields a little over
$57^{\prime\prime}$ per century! However, the full non-linear
theory of general relativity predicts the observed amount of
almost exactly $43^{\prime\prime}$ per century \footnote{The
actual observed perihelion precession of Mercury is approximately
$5600^{\prime\prime}$ per century. When effects such the
attraction due to the Newtonian gravitational field of the other
planets are taken into account, along with the fact that the
Earth is not an inertial frame of reference, a residual
$43^{\prime\prime}$ per century persists. The origin of this
residual precession was a mystery until general relativity was
formulated in 1915 \cite{ref:2}.} \cite{ref:2}.

\subsection{Perfect fluid in static gravity}
In this case, the equations (\ref{eq:6.1iv}), (\ref{eq:6.1v}) and (\ref{eq:6.4}) yield, for the metric tensor,
\begin{align}
g_{(0)ij}=&\left[\minusfactnabs\right]\delta_{ij}, \;\; g_{(0)i4}(x)\equiv 0, \;\; g_{(0)44}(x) =-\left[\plusfactnabs\right], \nonumber \\
g_{(0)}^{ij}=&\left[\minusfactnabs\right]^{-1}\delta^{ij}, \;\;  g_{(0)}^{i4}(x) \equiv 0, \;\; g_{(0)}^{44}(x)=- \left[\plusfactnabs\right]^{-1}. \label{eq:6.28}
\end{align}

\qquad The corresponding non-zero Christoffel symbols are provided by:
\begin{align}
\left\{
\begin{array}{l}
i \\
j \; k
\end{array}
\right\}_{(0)}=& \frac{1}{c^{2}} \left[\minusfactnabs\right]^{-1} \delta^{il} \left[ \delta_{jk} \frac{\partial W(\mathbf{x})}{\partial x^{l}} -\delta_{kl} \frac{\partial W(\mathbf{x})}{\partial x^{j}} -\delta_{lj} \frac{\partial W(\mathbf{x})}{\partial x^{k}} \right], \label{eq:6.29} \\
\left\{
\begin{array}{l}
4 \\
i \; 4
\end{array}
\right\}_{(0)} \equiv& \left\{
\begin{array}{l}
4 \\
4 \; i
\end{array}
\right\}_{(0)} =\frac{1}{c^{2}} \frac{\partial W(\mathbf{x})}{\partial x^{i}},\; \left\{
\begin{array}{l}
i \\
4 \; 4
\end{array}
\right\}_{(0)}= \frac{1}{c^{2}} \left[\minusfactnabs\right]^{-1} \delta^{ij} \frac{\partial W(\mathbf{x})}{\partial x^{j}} . \nonumber
\end{align}

\qquad The continuity equation (\ref{eq:5.30}) for a perfect fluid yields \cite{ref:15}
\begin{equation}
\frac{\partial}{\partial x^{b}} \left[\left(\rho+\frac{p}{c^{2}}\right) u^{b} \right] +\frac{\partial}{\partial x^{4}}\left[\left(\rho+\frac{p}{c^{2}}\right)u^{4}\right] =\frac{1}{c^{2}} g_{(0)\alpha\gamma}u^{\gamma} \left[\frac{\partial}{\partial x^{b}} \left(p g_{(0)}^{\alpha b}\right) +\frac{\partial}{\partial x^{4}} \left(p g_{(0)}^{\alpha 4} \right)\right]. \label{eq:6.30}
\end{equation}

\qquad Using (\ref{eq:6.29}), the equation (\ref{eq:6.30}) leads to
\begin{equation}
\frac{\partial}{\partial x^{b}} \left[\left(\rho+ \frac{p}{c^{2}}\right)u^{b}\right] +\frac{\partial}{\partial x^{4}} \left[\left(\rho +\frac{p}{c^{2}}\right) u^{4}\right] =\frac{1}{c^{2}} \left[\minusfactnabs\right]u^{b} \frac{\partial}{\partial x^{b}} \left[ p\left(\minusfactnabs\right)^{-1} \right]. \label{eq:6.31}
\end{equation}

\qquad Substituting (\ref{eq:6.9i}) and (\ref{eq:6.9ii}) into (\ref{eq:6.31}), we derive that
\begin{align}
&\frac{\partial}{\partial x^{b}} \left[ \left(\rho+ \frac{p}{c^{2}}\right) \frac{v^{b}}{\sqrt{\plusfactnabs -\left(\minusfactnabs\right)\frac{|\mathbf{v}|^{2}}{c^{2}}}} \right] +\frac{\partial}{\partial t} \left[ \frac{\left(\rho+\frac{p}{c^{2}}\right)}{\sqrt{\plusfactnabs -\left(\minusfactnabs\right)\frac{|\mathbf{v}|^{2}}{c^{2}}}}\right] \nonumber \\
=& \frac{1}{c^{2}} \left[\minusfactnabs \right] \frac{v^{b}}{\sqrt{ \plusfactnabs - \left(\minusfactnabs\right) \frac{|\mathbf{v}|^{2}}{c^{2}}}} \frac{\partial}{\partial x^{b}} \left[p\left(\minusfactnabs\right)^{-1}\right]. \label{eq:6.32}
\end{align}

\qquad Expanding the above in powers of $c^{-2}$, we can express
\begin{align}
& \frac{\partial}{\partial x^{b}} \left[\left(\rho+ \frac{p}{c^{2}}\right) \left(1-\frac{W(\mathbf{x})}{c^{2}} +\frac{|\mathbf{v}|^{2}}{2c^{2}}\right) v^{b}\right] +\frac{\partial}{\partial t} \left[\left(\rho+ \frac{p}{c^{2}}\right) \left(1-\frac{W(\mathbf{x})}{c^{2}} +\frac{|\mathbf{v}|^{2}}{2c^{2}}\right)\right] \nonumber \\
&= \frac{1}{c^{2}} v^{b} \frac{\partial p}{\partial x^{b}} +\mathcal{O}\left(\frac{1}{c^{4}}\right). \label{eq:6.33}
\end{align}
In more familiar notation the above equation reads
\begin{align}
&\mathbf{\nabla} \cdot \left[\rho \mathbf{v}\right] +\frac{\partial \rho}{\partial t} =\frac{1}{c^{2}} \left\{-p \mathbf{\nabla \cdot v} + \mathbf{\nabla}\cdot \left[\rho \left(W(\mathbf{x}) -\frac{|\mathbf{v}|^{2}}{2}\right) \mathbf{v}\right] \right. \nonumber \\
& \left. -\frac{\partial p}{\partial t} +\frac{\partial}{\partial t} \left[\rho\left(W(\mathbf{x}) -\frac{|\mathbf{v}|^{2}}{2}\right)\right] \right\} +\mathcal{O}\left(\frac{1}{c^{4}}\right).  \label{eq:6.34}
\end{align}
The relativistic correction terms are all collected on the right hand side of (\ref{eq:6.34}).

\qquad Now we shall investigate the equations of stream lines of a
perfect fluid in the external gravitational fluid. Equations
(\ref{eq:5.32}) provide
\begin{align}
&\left(\rho+\frac{p}{c^{2}}\right) \left[\frac{d^{2}\Chi^{i}(s)}{ds^{2}} +
\left\{
\begin{array}{l}
i \\
j \; k
\end{array}
\right\}_{(0)} \frac{d\Chi^{j}(s)}{ds}\frac{d\Chi^{k}(s)}{ds} +
\left\{
\begin{array}{l}
i \\
4 \; 4
\end{array}
\right\}_{(0)} \left(\frac{d\Chi^{4}(s)}{ds}\right)^{2} \right]_{|..} \nonumber \\
=&- \left[\delta^{i}_{\;\gamma} +\frac{1}{c^{2}} \left(g_{(0)\gamma j} \frac{d\Chi^{j}(s)}{ds} + g_{(0)\gamma 4} \frac{d\Chi^{4}(s)}{ds}\right) \frac{d\Chi^{i}}{ds}\right] \left[\frac{\partial}{\partial x^{b}} \left(p g_{(0)}^{b\gamma} \right)+ \frac{\partial}{\partial x^{4}}\left(pg_{(0)}^{4\gamma}\right)\right]_{|..} \nonumber \\
=& -\left[\delta^{i}_{\;k} +\frac{1}{c^{2}} g_{(0)kj} \frac{d\Chi^{i}}{ds}\frac{d\Chi^{j}}{ds} \right] \left[\frac{\partial}{\partial x^{b}} \left(p g_{(0)}^{bk}\right)\right]_{|..} -\frac{1}{c^{2}} g_{(0)44} \frac{d\Chi^{i}}{ds} \frac{d\Chi^{4}}{ds} \frac{\partial}{\partial x^{4}} \left(p g_{(0)}^{44}\right)_{|..}\, . \label{eq:6.35}
\end{align}
\qquad Using (\ref{eq:6.28}) and (\ref{eq:6.29}) along with (\ref{eq:6.35}), we deduce that
\begin{align}
& \left(\rho+ \frac{p}{c^{2}}\right) \left[ \frac{d^{2}\Chi^{i}(s)}{ds^{2}} + \frac{1}{c^{2}} \left(\minusfactnabs\right)^{-1} \left(\delta_{jk} \delta^{il} \frac{\partial W}{\partial x^{l}} -\delta^{i}_{\;k} \frac{\partial W}{\partial x^{j}} -\delta^{i}_{\;j} \frac{\partial W}{\partial x^{k}} \right)\frac{d\Chi^{j}(s)}{ds}\frac{d\Chi^{k}(s)}{ds}\right. \nonumber \\
&\left. + \frac{1}{c^{2}} \left(\minusfactnabs\right)^{-1} \delta^{ij} \frac{\partial W}{\partial x^{j}} \left(\frac{\partial \Chi^{4}(s)}{ds}\right)^{2} \right]_{|..} \nonumber \\
=&- \left[\delta^{i}_{\;k} +\frac{1}{c^{2}} \left(\minusfactnabs\right) \delta_{kj} \frac{d\Chi^{i}}{ds} \frac{d\Chi^{j}}{ds} \right] \left\{\frac{\partial}{\partial x^{a}} \left[p \left(\minusfactnabs\right)^{-1}
\delta^{ak} \right]\right\} \nonumber \\
&-\frac{1}{c^{2}} \left(\plusfactnabs\right)\frac{d\Chi^{i}}{ds} \frac{d\Chi^{4}}{ds} \left\{\frac{\partial}{\partial x^{4}} \left[p\left(\plusfactnabs\right)^{-1}\right] \right\}_{|..}\;\;\; . \label{eq:6.36}
\end{align}
We parameterize the curve by the rule:
\begin{equation}
\frac{d}{ds}=\left[\frac{d\overline{S}(t)}{dt}\right]^{-1}\frac{d}{dt}=
\frac{1}{\sqrt{\plusfactnabs-\left(\minusfactnabs\right)
\frac{|\mathbf{v}|^{2}}{c^{2}}}}_{\;|..}\, \frac{d}{dt}\;\; .
\label{eq:6.37}
\end{equation}
Then, from (\ref{eq:6.36}) and (\ref{eq:6.37}) the following is obtained:
\begin{align}
&\left(\rho +\frac{p}{c^{2}}\right) \left\{ \frac{dV^{i}(t)}{dt} -\frac{V^{i}(t)}{2} \frac{d}{dt} \left[\ln\left(\plusfactnabs -\left(\minusfactnabs\right)\frac{|\mathbf{v}|^{2}}{c^{2}}\right)\right] \right. \nonumber \\
&+ \frac{1}{c^{2}} \left(\minusfactnabs\right)^{-1} \left(\delta_{jk} \delta^{il} \frac{\partial W}{\partial x^{l}} -\delta^{i}_{\;k} \frac{\partial W}{\partial x^{j}} -\delta^{i}_{\;j} \frac{\partial W}{\partial x^{k}} \right) V^{j}V^{k} \nonumber \\
&+ \left.\left(\minusfactnabs\right)^{-1} \delta^{ij} \frac{\partial W}{\partial x^{j}} \right\}_{|..} \nonumber \\
=& -\left\{ \left[\plusfactnabs -\left(\minusfactnabs\right)\frac{|\mathbf{v}|^{2}}{c^{2}} \right] \delta^{i}_{\;k} +\frac{1}{c^{2}} \left(\minusfactnabs\right) \delta_{kj} V^{i}V^{j}\right\} \nonumber \\
&\times \left\{\delta^{ak} \frac{\partial}{\partial x^{a}}
\left[p\left(\minusfactnabs\right)^{-1}\right] \right\}_{|..}
-\frac{1}{c^{2}}\left(\plusfactnabs\right)V^{i}
\left\{\frac{\partial}{\partial t}
\left[p\left(\plusfactnabs\right)^{-1}\right]\right\}_{|..} .
\label{eq:6.38}
\end{align}

\qquad Expanding in inverse powers of $c^{2}$, we re-write the above equation of the stream-lines (\ref{eq:6.38}) as \cite{ref:15}:
\begin{align}
&\left\{\rho(\mathbf{x}) \left[\frac{d\mathbf{V}(t)}{dt} +\mathbf{\nabla}W(\mathbf{x})\right]
 +\mathbf{\nabla}p(x) \right\}_{|..} \nonumber \\
&=-\frac{1}{c^{2}} \left\{\rho \left[\left(\mathbf{V}(t)\cdot \frac{d\mathbf{V}(t)}{dt} -3 \mathbf{V}(t)\cdot \mathbf{\nabla}W\right) \mathbf{V}(t) + \left(2W +|\mathbf{V}(t)|^{2}\right)\mathbf{\nabla}W\right] \right. \nonumber \\
&+p \left. \left(\frac{d\mathbf{V}(t)}{dt} +3 \mathbf{\nabla}W\right) +\left(4 W -|\mathbf{V}(t)|^{2}]\right) \mathbf{\nabla}p + \left(\mathbf{V}(t)\cdot \mathbf{\nabla}p +\frac{\partial p}{\partial t} \right) \mathbf{V}(t) \right\}_{|..} +\mathcal{O}\left(\frac{1}{c^{4}}\right). \label{eq:6.39}
\end{align}
The right hand side contains all the relativistic corrections to the Euler equations of Newtonian fluid dynamics \cite{ref:12}, \cite{ref:15}.

\section{Generalizations to complicated materials and curvilinear coordinates}
\subsection{Perfect fluid plasma}
In this section we maintain field equations (\ref{eq:5.26i} - \ref{eq:5.26v}) for the fields $\phi_{(I)\alpha\beta}(x)$ and $\phi_{(0)\alpha\beta}(x)$. We also retain equations (\ref{eq:5.12i}) and (\ref{eq:5.12iii}) for the exterior metric $g_{(0)\mu\nu}(x)$ and the normalized vector $u^{\mu}$. Moreover, we stipulate the same equation as (\ref{eq:5.28}) for the interacting energy momentum stress tensor $\Theta^{\alpha\beta}(x)$.

\qquad A charged fluid or plasma satisfies the electromagnetic field equations (\ref{eq:4.10i}- \ref{eq:4.10iii}). The energy momentum stress tensor for this system is given by
\begin{equation}
T^{\alpha\beta}_{\mbox{\tiny (nvp)}}(x):= \left[
\rho(x)+\frac{p(x)}{c^{2}}\right]u^{\alpha}(x)u^{\beta}(x) +p(x)
g_{(0)}^{\alpha\beta}(x) +\Theta^{\alpha\beta}(x)
+\mathcal{M}^{\alpha\beta}(x) , \label{eq:7.1}
\end{equation}
where the subscript (nvp) denotes ``non-viscous plasma''. The
quantity $\mathcal{M}^{\alpha\beta}(x)$ is furnished by
(\ref{eq:4.12i}). Therefore, equation (\ref{eq:5.26iv}) implies
that
\begin{equation}
0=\frac{\partial T^{\alpha\beta}_{\mbox{\tiny
(nvp)}}(x)}{\partial x^{\beta}} = u^{\alpha}
\frac{\partial}{\partial x^{\beta}} \left[\left(\rho
+\frac{p}{c^{2}}\right) u^{\beta}\right]
+\left(\rho+\frac{p}{c^{2}}\right) \nabla^{(0)}_{\beta}
u^{\alpha} +\frac{\partial}{\partial x^{\beta}}\left(p
g^{\alpha\beta}_{(0)}\right) +\frac{J^{\beta}}{c}
F_{\beta}^{\;\;\alpha}. \label{eq:7.2}
\end{equation}
Here, we have used equation (\ref{eq:5.28}) for $\frac{\partial
\Theta^{\alpha\beta}(x)}{\partial x^{\beta}}$.

\qquad Multiplying (\ref{eq:7.2}) by $\overline{u}_{\alpha}$ (and
summing), we derive the plasma continuity equation:
\begin{equation}
\frac{\partial}{\partial x^{\beta}} \left[\left(\rho +
\frac{p}{c^{2}} \right)u^{\beta} \right]
=\frac{\overline{u}_{\alpha}}{c^{2}}
\left[\frac{\partial}{\partial x^{\beta}} \left(p
g_{(0)}^{\alpha\beta}\right) +\frac{J^{\beta}}{c}
F_{\beta}^{\;\;\alpha} \right]. \label{eq:7.3}
\end{equation}
Substituting this last equation into (\ref{eq:7.2}), we deduce the
generalized Euler equation
\begin{equation}
\left( \rho+\frac{p}{c^{2}}\right) u^{\beta} \nabla^{(0)}_{\beta} u^{\alpha} +\left(\delta^{\alpha}_{\;\gamma} + \frac{\overline{u}_{\gamma}u^{\alpha}}{c^{2}}\right) \left[ \frac{\partial}{\partial x^{\beta}} \left(p g_{(0)}^{\gamma\beta}\right) + \frac{J^{\beta}}{c} F_{\beta}^{\;\;\gamma}\right]=0. \label{eq:7.4}
\end{equation}

\qquad From (\ref{eq:7.4}), the equations for a stream line emerges as
\begin{align}
& \left\{ \left[\rho +\frac{p}{c^{2}}\right] \frac{d^{2}\Chi^{\alpha}(s)}{ds^{2}} +
\left\{
\begin{array}{l}
\alpha  \\
\beta \; \gamma
\end{array}
\right\}_{(0)}
 \frac{d\Chi^{\beta}(s)}{ds} \frac{d\Chi^{\gamma}(s)}{ds} \right\}_{|x^{\alpha}=\Chi^{\alpha}(s)} \nonumber \\
&= -\left\{ \left[ \delta^{\alpha}_{\;\gamma} +
\frac{g_{(0)\mu\gamma}(x)}{c^{2}}  \frac{d\Chi^{\mu}(s)}{ds}
\frac{d\Chi^{\alpha}(s)}{ds} \right] \left[
\frac{\partial}{\partial x^{\beta}} \left(p(x)
g_{(0)}^{\gamma\beta}(x)\right) +\frac{J^{\beta}(x)}{c}
F_{\beta}^{\;\;\gamma}(x) \right] \right\}_{|..}\; .
\label{eq:7.5}
\end{align}

\qquad One may add viscosity to the above system. For such a fluid, the energy momentum stress tensor is
\begin{align}
T^{\alpha\beta}(x):= &T^{\alpha\beta}_{\mbox{\tiny(nvp)}} - \eta(x) \left\{ \left[g_{(0)}^{\alpha\sigma}(x) +\frac{u^{\alpha}(x)u^{\sigma}(x)}{c^{2}}\right] \frac{\partial u^{\beta}(x)}{\partial x^{\sigma}} +\left[g_{(0)}^{\beta\sigma}(x) +\frac{u^{\beta}(x)u^{\sigma}(x)}{c^{2}}\right]\frac{\partial u^{\alpha}(x)}{\partial x^{\sigma}} \right\} \nonumber \\
&+ \left[\frac{2}{3} \eta(x) -\zeta(x) \right] \left[g_{(0)}^{\alpha\beta}(x) + \frac{u^{\alpha}(x) u^{\beta}(x)}{c^{2}}\right] \frac{\partial u^{\sigma}(x)}{\partial x^{\sigma}} . \label{eq:7.6}
\end{align}
Here, $\eta(x)$ and $\zeta(x)$ represent the \emph{shear viscosity} and \emph{bulk viscosity} respectively. Using (\ref{eq:5.26iv}), we obtain, from (\ref{eq:7.6}) the following equation
\begin{align}
0= \frac{\partial T^{\alpha\beta}(x)}{\partial x^{\beta}} = &\frac{\partial T^{\alpha\beta}_{\mbox{\tiny{(nvp)}}}(x)}{\partial x^{\beta}} -\frac{\partial}{\partial x^{\beta}} \left\{ \eta \left[\left(g_{(0)}^{\alpha\sigma} +\frac{u^{\alpha}u^{\sigma}}{c^{2}} \right) \frac{\partial u^{\beta}}{\partial x^{\sigma}} + \left(g_{(0)}^{\beta\sigma} +\frac{u^{\beta}u^{\sigma}}{c^{2}} \right) \frac{\partial u^{\alpha}}{\partial x^{\sigma}} \right] \right. \nonumber \\
&+\left. \left(\zeta -\frac{2}{3}\eta\right) \left(g_{(0)}^{\alpha\beta} +\frac{u^{\alpha}u^{\beta}}{c^{2}} \right) \frac{\partial u^{\sigma}}{\partial x^{\sigma}} \right\}. \label{eq:7.7}
\end{align}
Here we have again stipulated the equation (\ref{eq:5.28}).

\qquad Multiplying (\ref{eq:7.7}) by $\overline{u}_{\alpha}$, we derive the continuity equation:
\begin{align}
\frac{\partial}{\partial x^{\beta}} \left[\left(\rho + \frac{p}{c^{2}} \right)u^{\beta} \right] =& \frac{\overline{u}_{\alpha}}{c^{2}} \left\{\left[\frac{\partial}{\partial x^{\beta}} \left(p g_{(0)}^{\alpha\beta}\right) +\frac{J^{\beta}}{c} F_{\beta}^{\;\;\alpha} \right] \right. \nonumber \\
& - \frac{\partial}{\partial x^{\beta}} \left(\eta \left[ \left( g_{(0)}^{\alpha\sigma} +\frac{u^{\alpha}u^{\sigma}}{c^{2}} \right) \frac{\partial u^{\beta}}{\partial x^{\sigma}} + \left( g_{(0)}^{\beta\sigma} +\frac{u^{\beta}u^{\sigma}}{c^{2}} \right) \frac{\partial u^{\alpha}}{\partial x^{\sigma}} \right] \right. \nonumber \\
& +\left. \left. \left(\zeta-\frac{2}{3}\eta\right) \left( g_{(0)}^{\alpha\beta} +\frac{u^{\alpha}u^{\beta}}{c^{2}} \right) \frac{\partial u^{\sigma}}{\partial x^{\sigma}} \right) \right\} \label{eq:7.8}
\end{align}
Substituting (\ref{eq:7.8}) into (\ref{eq:7.7}), we deduce the generalized \emph{Navier-Stokes equation}:
\begin{align}
\left( \rho+\frac{p}{c^{2}}\right) u^{\beta} \nabla^{(0)}_{\beta} u^{\alpha} =& \left(\delta^{\alpha}_{\;\gamma} + \frac{\overline{u}_{\gamma}u^{\alpha}}{c^{2}}\right) \left\{ -\frac{\partial}{\partial x^{\beta}} \left(p g_{(0)}^{\gamma\beta}\right) - \frac{J^{\beta}}{c} F_{\beta}^{\;\;\gamma} \right. \nonumber \\
&+ \frac{\partial}{\partial x^{\beta}} \left( \eta \left[ \left(g_{(0)}^{\gamma\sigma} +\frac{u^{\gamma}u^{\sigma}}{c^{2}} \right) \frac{\partial u^{\beta}}{\partial x^{\sigma}} +\left(g_{(0)}^{\beta\sigma} +\frac{u^{\beta}u^{\sigma}}{c^{2}} \right) \frac{\partial u^{\gamma}}{\partial x^{\sigma}}\right] \right.  \nonumber \\
&\left. \left. + \left(\zeta-\frac{2}{3}\eta\right) \left( g_{(0)}^{\gamma\beta} +\frac{u^{\gamma}u^{\beta}}{c^{2}}\right) \frac{\partial u^{\sigma}}{\partial x^{\sigma}} \right) \right\} . \label{eq:7.9}
\end{align}

\subsection{Curvilinear coordinates and orthonormal frames}
Now we shall introduce curvilinear spacetime coordinates by transformation equations:
\begin{align}
\hat{x}^{\alpha}=&\hat{X}^{\alpha}(x), \nonumber \\
\frac{\partial \left(\hat{x}^{1}, \hat{x}^{2},\hat{x}^{3}, \hat{x}^{4} \right)}{\partial
(x^{1}, x^{2}, x^{3}, x^{4})} & \neq 0, \label{eq:7.10} \\
x^{\alpha}=& X^{\alpha}(\hat{x}). \nonumber
\end{align}
Here, we have assumed that the functions $\hat{X}^{\alpha}$ are of class $C^{2}$. The transformation rules for tensor fields are furnished by \cite{ref:7}
\begin{equation}
\hat{T}^{\alpha\beta ..}_{\;\;\;\;\; \mu\nu ..}(\hat{x})= \frac{\partial \hat{X}^{\alpha}(x)}{\partial x^{\gamma}} \frac{\partial \hat{X}^{\beta}(x)}{\partial x^{\delta}} ..
\frac{\partial X^{\rho}(\hat{x})}{\partial \hat{x}^{\mu}} \frac{\partial X^{\sigma}(\hat{x})}{\partial \hat{x}^{\nu}}\, T^{\gamma\delta..}_{\;\;\;\;\; \rho \sigma ..}(x). \label{eq:7.11}
\end{equation}

\qquad The coordinate transformation (\ref{eq:7.10}) generate a \emph{new} metric tensor field as follows:
\begin{align}
\hat{g}_{\alpha\beta}(\hat{x}):=& d_{\mu\nu} \frac{\partial X^{\mu}(\hat{x})}{\partial \hat{x}^{\alpha}} \frac{\partial X^{\nu}(\hat{x})}{\partial \hat{x}^{\beta}} , \nonumber \\
\left[ \hat{g}^{\alpha\beta}(\hat{x}) \right]:=&\left[\hat{g}_{\alpha\beta}(\hat{x})\right]^{-1} . \label{eq:7.12}
\end{align}
The Christoffel symbols are given by (compare with (\ref{eq:5.6}))
\begin{equation}
\widehat{\left\{
\begin{array}{l}
\alpha \\
\beta \; \gamma
\end{array}
\right\}}:=\frac{1}{2} \hat{g}^{\alpha\sigma}(\hat{x}) \left[ \frac{\partial \hat{g}_{\gamma\sigma}(\hat{x})}{\partial \hat{x}^{\beta}} + \frac{\partial \hat{g}_{\sigma\beta}(\hat{x})}{\partial \hat{x}^{\gamma}} - \frac{\partial \hat{g}_{\beta\gamma}(\hat{x})}{\partial \hat{x}^{\sigma}} \right] . \label{eq:7.13}
\end{equation}
The covariant derivatives are defined by \cite{ref:7}, \cite{ref:19} (compare with (\ref{eq:5.10i} - \ref{eq:5.10iv}))
\begin{align}
\hat{\nabla}_{\alpha} \hat{T}^{\mu\nu ..}_{\;\;\;\;\; \rho\sigma..} (\hat{x}) :=&\frac{\partial \hat{T}^{\mu\nu ..}_{\;\;\;\;\;\rho\sigma..}(\hat{x})}{\partial \hat{x}^{\alpha}} +
\widehat{\left\{
\begin{array}{l}
\mu \\
\alpha \; \beta
\end{array}
\right\}}
\hat{T}^{\beta\nu ..}_{\;\;\;\;\; \rho\sigma..} (\hat{x}) +
\widehat{\left\{
\begin{array}{l}
\nu \\
\alpha \; \beta
\end{array}
\right\}}
\hat{T}^{\mu\beta ..}_{\;\;\;\;\; \rho\sigma..} (\hat{x}) + ..  \nonumber \\
&- \widehat{\left\{
\begin{array}{l}
\beta \\
\alpha \; \rho
\end{array}
\right\}}
\hat{T}^{\mu\nu ..}_{\;\;\;\;\; \beta\sigma..} (\hat{x}) -
\widehat{\left\{
\begin{array}{l}
\beta \\
\alpha \; \sigma
\end{array}
\right\}}
\hat{T}^{\mu\nu ..}_{\;\;\;\;\; \rho\beta..} (\hat{x}) -.. \;\;\;. \label{eq:7.14}
\end{align}

\qquad If we replace the various tensor fields $T^{\gamma\delta ..}_{\;\;\;\;\;\rho\sigma ..}(x)$ by $\hat{T}^{\alpha\beta ..}_{\;\;\;\;\;\mu\nu..}(\hat{x})$ (in (\ref{eq:7.11})), and $\hat{\nabla}_{\alpha} \hat{T}^{\mu\nu ..}_{\;\;\;\;\;\rho\sigma}(\hat{x})$ (in (\ref{eq:7.14}), then all the constitutive equations are expressed in curvilinear spacetime coordinates $\hat{x}^{\alpha}$.

\qquad However, for the sake of applications, we restrict ourselves to \emph{spatial curvilinear coordinates} only. In that case the following restriction is placed on equations (\ref{eq:7.10})
\begin{equation}
\hat{x}^{i}= \hat{X}^{i}(\mathbf{x}), \;\;\hat{x}^{4}=\hat{X}^{4}(\mathbf{x},x^{4}):=x^{4}, \;\; {x}^{i}= {X}^{i}(\mathbf{x}). \label{eq:7.15}
\end{equation}
The tensor transformation rules (\ref{eq:7.11}) then simplify in the obvious way:
\begin{align}
\hat{T}^{ij4..}_{\;\;\;\;\;\;\;kl4..}(\hat{\mathbf{x}},\hat{x}^{4})=& \frac{\partial \hat{X}^{i}(\mathbf{x})}{\partial x^{a}} \frac{\partial \hat{X}^{j}(\mathbf{x})}{\partial x^{b}}.. \frac{\partial {X}^{m}(\hat{\mathbf{x}})}{\partial \hat{x}^{k}} \frac{\partial {X}^{n}(\hat{\mathbf{x}})}{\partial \hat{x}^{l}}.. T^{ab4..}_{\;\;\;\;\;\;\;mn4..}(\mathbf{x},x^{4}), \nonumber \\
\hat{T}^{4}_{\;4}(\hat{\mathbf{x}}, \hat{x}^{4}) =& T^{4}_{\;4}(\mathbf{x}, x^{4}),  \label{eq:7.16}
\end{align}
and the metric tensor field in (\ref{eq:7.2}) reduces to
\begin{align}
\hat{g}_{ij}(\hat{\mathbf{x}})=& \delta_{kl} \frac{\partial X^{k}(\hat{\mathbf{x}})}{\partial \hat{x}^{i}} \frac{\partial X^{l}(\hat{\mathbf{x}})}{\partial \hat{x}^{j}} , \nonumber \\
\hat{g}_{i4}(\hat{\mathbf{x}},x^{4})\equiv& 0, \label{eq:7.17} \\
\hat{g}_{44}(\hat{\mathbf{x}},\hat{x}^{4})\equiv&  -1. \nonumber
\end{align}

\qquad The Christoffel symbols in (\ref{eq:7.13}) boil down to
\begin{align}
\widehat{\left\{
\begin{array}{l}
i \\
j \; k
\end{array}
\right\}} =&\frac{1}{2} \hat{g}^{ia}(\hat{\mathbf{x}}) \left[ \frac{\partial \hat{g}_{ka}(\hat{\mathbf{x}})}{\partial \hat{x}^{j}} + \frac{\partial \hat{g}_{aj}(\hat{\mathbf{x}})}{\partial \hat{x}^{k}} - \frac{\partial \hat{g}_{jk}(\hat{\mathbf{x}})}{\partial \hat{x}^{a}} \right], \nonumber \\
\widehat{\left\{
\begin{array}{l}
4 \\
j \; k
\end{array}
\right\}} =&
\widehat{\left\{
\begin{array}{l}
4 \\
j \; 4
\end{array}
\right\}} =
\widehat{\left\{
\begin{array}{l}
4 \\
4 \; 4
\end{array}
\right\}} \equiv 0. \label{eq:7.18}
\end{align}

\qquad The laws of covariant differentiation (\ref{eq:7.14}) imply that
\begin{align}
\hat{\nabla}_{j}\hat{T}^{a4}_{\;\;\;\;\;b4}(\hat{\mathbf{x}}, x^{4}) =& \frac{\partial \hat{T}^{a4}_{\;\;\;\;\;b4}(\hat{\mathbf{x}}, \hat{x}^{4})}{\partial \hat{x}^{j}} +
\widehat{\left\{
\begin{array}{l}
a \\
j \; k
\end{array}
\right\}}
\hat{T}^{k4}_{\;\;\;\;\;b4} (\hat{\mathbf{x}}, \hat{x}^{4}) -
\widehat{\left\{
\begin{array}{l}
k \\
j \; b
\end{array}
\right\}}
\hat{T}^{a4}_{\;\;\;\;\;k4} (\hat{\mathbf{x}}, \hat{x}^{4}),
\nonumber \\
\hat{\nabla}_{4}\hat{T}^{a4}_{\;\;\;\;\;b4}(\hat{\mathbf{x}},
x^{4}) =& \frac{\partial
\hat{T}^{a4}_{\;\;\;\;\;b4}(\hat{\mathbf{x}},
\hat{x}^{4})}{\partial \hat{x}^{4}} = \frac{\partial
\hat{T}^{a4}_{\;\;\;\;\;b4}(\hat{\mathbf{x}}, {x}^{4})}{\partial
{x}^{4}}. \label{eq:7.19}
\end{align}

\qquad If we now replace various tensor fields
$T^{\mu\nu..}_{\;\;\;\;\; \alpha\beta ..}$,
$\frac{\partial}{\partial x^{\gamma}}T^{\mu\nu..}_{\;\;\;\;\;
\alpha\beta ..}$ occurring in the constitutive equations
(\ref{eq:7.1} - \ref{eq:7.9}) by $\hat{T}^{ab..}_{\;\;\;\;\; cd
..}$, $\hat{T}^{a4..}_{\;\;\;\;\; c4 ..}$,
$\hat{T}^{44..}_{\;\;\;\;\; 44 ..}$ and
$\hat{\nabla}_{j}\hat{T}^{ab..}_{\;\;\;\;\;cd ..}
(\hat{\mathbf{x}}, \hat{x}^{4})$, $\frac{\partial}{\partial
x^{4}} \hat{T}^{ab..}_{\;\;\;\;\;cd...}(\hat{\mathbf{x}},
\hat{x}^{4})$ etc., then we have converted all relevant equations
into spatial curvilinear coordinates.

\qquad In physical applications, usually \emph{orthonormal} or
\emph{physical} components of a tensor are necessary
\cite{ref:19}. For that purpose we introduce three orthonormal
vectors, $\vec{\lambda}_{A}(\mathbf{x})$ $A \in
\left\{1,2,3\right\}$, in space. These vectors satisfy the
orthonormality conditions:
\begin{equation}
\hat{g}_{ij}(\hat{\mathbf{x}}) \lambda^{i}_{A}(\hat{\mathbf{x}})
\lambda^{j}_{B}(\hat{\mathbf{x}}) =\delta_{AB} .\label{eq:7.20}
\end{equation}

\qquad We define the inverse entries by
\begin{align}
\left[\mu^{A}_{i}\right]:=& \left[\lambda^{i}_{A}\right]^{-1}, \nonumber \\
\lambda^{i}_{A}\mu^{A}_{j}=& \delta^{i}_{\;j},  \label{eq:7.21} \\
\mu^{A}_{i}\lambda^{j}_{A}=&\delta^{j}_{\;i} . \nonumber
\end{align}
(Here, \emph{the summation convention is also followed for capital roman indices}.)

\qquad By (\ref{eq:7.20}) and (\ref{eq:7.21}) we obtain
\begin{equation}
 \hat{g}_{ij}(\hat{\mathbf{x}})=\delta_{AB} \mu^{A}_{\;i}(\hat{\mathbf{x}}) \mu^{B}_{\; j}(\hat{\mathbf{x}}) . \label{eq:7.22}
\end{equation}

\qquad The tensor transformation rules (\ref{eq:7.11}) lead to
\begin{equation}
\hat{T}^{AB..}_{\;\;\;\;\;CD..}(\hat{\mathbf{x}}, \hat{x}^{4}) =
\mu^{A}_{a}(\hat{\mathbf{x}}) \mu^{B}_{b}(\hat{\mathbf{x}})..
\lambda^{i}_{C}(\hat{\mathbf{x}})
\lambda^{j}_{D}(\hat{\mathbf{x}})..
\hat{T}^{ab..}_{\;\;\;\;\;ij..}(\hat{\mathbf{x}}, \hat{x^{4}}).
\label{eq:7.23}
\end{equation}

\qquad Instead of Christoffel symbols, we require \emph{Ricci
rotation coefficients} \cite{ref:19} for the connexion. These are
defined by:
\begin{align}
\gamma_{ABC}(\hat{\mathbf{x}}):=& \hat{g}_{jl}(\hat{\mathbf{x}}) \left(\hat{\nabla}_{k} \lambda^{l}_{A} \right) \lambda^{j}_{B}(\hat{\mathbf{x}})\lambda^{k}_{C}(\hat{\mathbf{x}}) \equiv -\gamma_{BAC}(\hat{\mathbf{x}}), \nonumber \\
\gamma^{A}_{\;\;BC}(\hat{\mathbf{x}}):=& \delta^{AJ}
\gamma_{JBC}(\hat{\mathbf{x}}) =\gamma_{ABC}(\hat{\mathbf{x}}).
\label{eq:7.24}
\end{align}

\qquad The appropriate covariant derivatives can be characterized by:
\begin{equation}
\hat{\nabla}_{J} \hat{T}^{A4}_{\;\;\;\;\;B4}(\hat{\mathbf{x}},\hat{x}^{4}):= \lambda^{i}_{J}(\mathbf{x}) \frac{\partial}{\partial \hat{x}^{i}} \hat{T}^{A4}_{\;\;\;\;\;B4}(\hat{\mathbf{x}}, \hat{x}^{4}) -\gamma^{A}_{\;\;DJ}(\hat{\mathbf{x}}) \hat{T}^{D4}_{\;\;\;\;\;B4} + \gamma^{D}_{\;\;BJ}(\hat{\mathbf{x}}) \hat{T}^{A4}_{\;\;\;\;\;D4}. \label{eq:7.25}
\end{equation}

\qquad If we now replace the tensor fields $T^{\mu\nu..}_{\;\;\;\;\;\alpha\beta}(x)$, $\frac{\partial}{\partial x^{\gamma}} T^{\mu\nu..}_{\;\;\;\;\;\alpha\beta}(x)$ appearing in the constitutive equations (\ref{eq:7.1} - \ref{eq:7.9}) by $\hat{T}^{AB..}_{\;\;\;\;\;CD..}(\hat{\mathbf{x}}, \hat{x}^{4})$, $\hat{T}^{A4..}_{\;\;\;\;\;C4..}(\hat{\mathbf{x}}, \hat{x}^{4})$, $\hat{\nabla}_{J}\hat{T}^{AB..}_{\;\;\;\;\;CD..}(\hat{\mathbf{x}}, \hat{x}^{4})$, $\frac{\partial}{\partial \hat{x}^{4}} \hat{T}^{AB..}_{\;\;\;\;\;CD..}(\hat{\mathbf{x}}, \hat{x}^{4})$ etc., then we have transformed all the required equations into the orthonormal or physical frame. Physical measurements correspond to quantities expressed in this frame.

\qquad Before concluding, we shall now explore a special example
which is most useful for continuum mechanics. This example
involves orthogonal coordinate systems in Euclidean three-space.
The equations (\ref{eq:7.17}) reduce to
\begin{equation}
\left[\hat{g}_{ij}(\hat{\mathbf{x}})\right] = \left[
\begin{array}{ccc}
\left[h_{1}(\hat{\mathbf{x}})\right]^{2} & 0 & 0 \\
0 & \left[h_{2}(\hat{\mathbf{x}})\right]^{2}& 0 \\
0 & 0 & \left[h_{3}(\hat{\mathbf{x}})\right]^{2}
\end{array}
\right], \label{eq:7.26}
\end{equation}
\begin{equation}
h_{i}(\mathbf{x}) > 0, \;\; \sqrt{\mbox{det}\left[\hat{g}_{ij}\right]} = h_{1}(\hat{\mathbf{x}})h_{2}(\hat{\mathbf{x}})h_{3}(\hat{\mathbf{x}}) >0 . \nonumber
\end{equation}

\qquad The non-zero Christoffel symbols, from (\ref{eq:7.18}) and
(\ref{eq:7.26}) are summarized by:
\begin{align}
\widehat{\left\{
\begin{array}{l}
1 \\
1 \; 1
\end{array}
\right\}}= \frac{\partial}{\partial \hat{x}^{1}} \ln h_{1}, \;\; \mbox{etc. }; \nonumber \\
\widehat{\left\{
\begin{array}{l}
1 \\
1 \; 2
\end{array}
\right\}}= \frac{\partial}{\partial \hat{x}^{2}} \ln h_{1}, \;\; \mbox{etc. }; \label{eq:7.27}\\
\widehat{\left\{
\begin{array}{l}
1 \\
2 \; 2
\end{array}
\right\}}= - h_{2} \left[h_{1}\right]^{-2}
\frac{\partial}{\partial \hat{x}^{1}} \ln h_{2}, \;\; \mbox{etc.}
\nonumber
\end{align}

\qquad The orthonormal (or physical) vector components from (\ref{eq:7.20}) and (\ref{eq:7.26}) are given by
\begin{equation}
\lambda^{i}_{\;A}(\hat{\mathbf{x}}) =\left[h_{i}(\hat{\mathbf{x}})\right]^{-1} \delta^{i}_{\;A}, \;\; \mu^{A}_{\;i}(\hat{\mathbf{x}}) =h_{i}(\hat{\mathbf{x}}) \delta^{A}_{\;i} \; ,\label{eq:7.28}
\end{equation}
and the non-zero Ricci rotation coefficients from (\ref{eq:7.24}), (\ref{eq:7.27}) and (\ref{eq:7.28}) are furnished by:
\begin{align}
\gamma_{ABC}:=&\delta_{AE}\gamma^{E}_{\;BC},\;\;\;\gamma_{(A)(B)(C)}:=\gamma_{ABC} \equiv \gamma_{BAC}, \nonumber \\
\gamma_{(2)(1)(2)}=& -\frac{\partial}{\partial \hat{x}^{1}} \ln h_{2}, \;\; \gamma_{(3)(1)(3)}= -\frac{\partial}{\partial \hat{x}^{1}} \ln h_{3}\; , \nonumber \\
\gamma_{(1)(2)(1)}=& -\frac{\partial}{\partial \hat{x}^{2}} \ln h_{1}, \;\; \gamma_{(3)(2)(3)}= -\frac{\partial}{\partial \hat{x}^{2}} \ln h_{3} \;,  \label{eq:7.29} \\
\gamma_{(1)(3)(1)}=& -\frac{\partial}{\partial \hat{x}^{3}} \ln h_{1}, \;\; \gamma_{(2)(3)(2)}= -\frac{\partial}{\partial \hat{x}^{3}} \ln h_{2}. \nonumber
\end{align}

\qquad Also, the gradient of a scalar field is given by
\begin{align}
\hat{\nabla}_{i} \hat{\phi}(\hat{\mathbf{x}})=& \frac{\partial}{\partial \hat{x}^{i}} \hat{\phi}(\hat{\mathbf{x}}), \; \hat{\nabla}_{A} \hat{\phi}(\hat{\mathbf{x}}) = \lambda^{i}_{\; A} \frac{\partial}{\partial \hat{x}^{i}} \hat{\phi}(\hat{\mathbf{x}}) = \left(h_{A}\right)^{-1} \frac{\partial}{\partial \hat{x}^{A}} \hat{\phi}(\hat{\mathbf{x}}), \nonumber \\
\hat{\nabla}_{(1)}\hat{\phi}(\hat{\mathbf{x}})=& (h_{1})^{-1} \frac{\partial}{\partial \hat{x}^{1}} \hat{\phi}(\hat{\mathbf{x}}) \not\equiv \hat{\nabla}_{1} \hat{\phi}(\hat{\mathbf{x}}). \label{eq:7.30}
\end{align}
In (\ref{eq:7.30}) there is no summation over $A$.

\qquad The divergence of a vector field is furnished by
\begin{equation}
\hat{\nabla}_{i} \hat{T}^{i}(\hat{\mathbf{x}}) \equiv
\hat{\nabla}_{A} \hat{T}^{A}(\hat{\mathbf{x}})
=\left(h_{1}h_{2}h_{3}\right)^{-1} \left\{
\frac{\partial}{\partial \hat{x}^{i}}
\left[\left(h_{1}h_{2}h_{3}\right) \hat{T}^{i}(\hat{\mathbf{x}})
\right]\right\}. \label{eq:7.31}
\end{equation}
The curl is given by:
\begin{align}
\left[\hat{\nabla}\times\hat{\mathbf{A}}\right]^{i}:=& \frac{1}{2}
\frac{\epsilon^{ijk}}{\sqrt{\mbox{det}[\hat{g}_{ab}]}} \left[
\hat{\nabla}_{j}\hat{A}_{k} - \hat{\nabla}_{k} \hat{A}_{j}
\right] =
\frac{1}{2(h_{1}h_{2}h_{3})} \epsilon^{ijk} \left[\frac{\partial \hat{A}_{k}(\hat{\mathbf{x}})}{\partial \hat{\mathbf{x}}^{j}} - \frac{\partial \hat{A}_{j}(\hat{\mathbf{x}})}{\partial \hat{\mathbf{x}}^{k}}\right], \nonumber \\
\left[\hat{\nabla}\times\hat{\mathbf{A}}\right]^{B}:=& \frac{1}{2} \epsilon^{BCD} \left[\lambda^{i}_{\; C} \frac{\partial}{\partial \hat{x}^{i}} \left(\hat{A}_{D}(\hat{\mathbf{x}})\right) - \lambda^{i}_{\; D}\frac{\partial}{\partial \hat{x}^{i}} \left(\hat{A}_{C}(\hat{\mathbf{x}})\right)\right] \nonumber \\
=&\frac{1}{2} \epsilon^{BCD} \left[(h_{C})^{-1} \frac{\partial}{\partial \hat{x}^{C}} \left(\hat{A}_{D}(\hat{\mathbf{x}})\right) - (h_{D})^{-1}\frac{\partial}{\partial \hat{x}^{D}} \left(\hat{A}_{C}(\hat{\mathbf{x}})\right)\right], \label{eq:7.32} \\
\left[\hat{\nabla}\times\hat{\mathbf{A}}\right]^{1} \not\equiv&
\left[\hat{\nabla}\times\hat{\mathbf{A}}\right]^{(1)}. \nonumber
\end{align}

\qquad Finally, the Laplacian operator is furnished by
\begin{align}
&\nabla^{2}\hat{W}(\hat{\mathbf{x}}):= \hat{g}^{ij} \hat{\nabla}_{i} \hat{\nabla}_{j}\hat{W}(\hat{\mathbf{x}})\equiv \delta^{AB} \hat{\nabla}_{A} \hat{\nabla}_{B}\hat{W}(\hat{\mathbf{x}}) \label{eq:7.33} \\
&= \frac{1}{\left(h_{1}h_{2}h_{3}\right)} \left[ \frac{\partial}{\partial \hat{x}^{1}} \left( \frac{h_{1}h_{3}}{h_{1}} \frac{\partial \hat{W}(\hat{\mathbf{x}})}{\partial \hat{x}^{1}} \right) + \frac{\partial}{\partial \hat{x}^{2}} \left( \frac{h_{3}h_{1}}{h_{2}} \frac{\partial \hat{W}(\hat{\mathbf{x}})}{\partial \hat{x}^{2}} \right) + \frac{\partial}{\partial \hat{x}^{3}} \left( \frac{h_{1}h_{2}}{h_{3}} \frac{\partial \hat{W}(\hat{\mathbf{x}})}{\partial \hat{x}^{3}} \right) \right]. \nonumber
\end{align}

\newpage
\linespread{0.6}
\bibliographystyle{unsrt}

\end{document}